\documentclass[a4paper]{article}
 \pdfoutput=1
\usepackage{mathtools}
\mathtoolsset{showonlyrefs=true}

\usepackage{algorithm}
\usepackage{algorithmic}
\usepackage{hyperref}
\usepackage{url}

\usepackage{longtable}
\usepackage{ifthen}
\newboolean{IEEE}
\setboolean{IEEE}{false}
\newboolean{ACM}
\setboolean{ACM}{false}

\usepackage[section]{placeins}
\begin{document}

\title{Precise Energy Modeling for the Bluetooth Low Energy Protocol}
\author{
PHILIPP KINDT, DANIEL YUNGE, ROBERT DIEMER \\and SAMARJIT CHAKRABORTY,\\ Institute for Real-Time Computer Systems, Technische Universit\"at M\"unchen
}

\input


\ifACM
\begin{abstract}
 Bluetooth Low Energy (BLE) is a wireless protocol well suited for ultra-low-power sensors running on small batteries. It is optimized for low power communication and is not compatible with the original Bluetooth, referred to as Bluetooth Basic Rate (BR)/Enhanced Data Rate (EDR). BLE is described as a new protocol in the official Bluetooth 4.0 specification.
 It reuses many parts of the Bluetooth BR hardware and software stack to enable dual mode devices supporting Bluetooth BR/EDR and BLE. 
 To design energy-efficient devices, the protocol provides a number of parameters that need to be optimized within an energy, latency and throughput design space.
 To minimize power consumption, the protocol parameters have to be optimized for a given application. Therefore, an energy-model that can predict the 
 energy consumption of a BLE-based wireless device for different parameter value settings, is needed. As BLE differs from \mbox{Bluetooth BR} significantly, models for Bluetooth BR cannot be easily applied to the BLE protocol. Since the last one year, there have been a couple of proposals on energy models for BLE. However, none of them can model all the operating modes of the protocol. This paper presents a precise energy model of the BLE protocol, that allows the computation of a device's power consumption in all possible operating modes. To the best of our knowledge, our proposed model is not only one of the most accurate ones known so far (because it accounts for all protocol parameters), but it is also the only one that models all the operating modes of BLE. Furthermore, we present a sensitivity analysis of the different parameters on the energy consumption and evaluate the accuracy of the model using both discrete event simulation and actual measurements. Based on this model, guidelines for system designers are presented, that help choosing the right parameters for optimizing the energy consumption for a given application.
\end{abstract}

%
\maketitle
\else
\maketitle
\begin{abstract}
 Bluetooth Low Energy (BLE) is a wireless protocol well suited for ultra-low-power sensors running on small batteries. It is optimized for low power communication and is not compatible with the original Bluetooth, referred to as Bluetooth Basic Rate (BR)/Enhanced Data Rate (EDR). BLE is described as a new protocol in the official Bluetooth 4.0 specification.
 It reuses many parts of the Bluetooth BR hardware and software stack to enable dual mode devices supporting Bluetooth BR/EDR and BLE. 
 To design energy-efficient devices, the protocol provides a number of parameters that need to be optimized within an energy, latency and throughput design space.
 To minimize power consumption, the protocol parameters have to be optimized for a given application. Therefore, an energy-model that can predict the 
 energy consumption of a BLE-based wireless device for different parameter value settings, is needed. As BLE differs from \mbox{Bluetooth BR} significantly, models for Bluetooth BR cannot be easily applied to the BLE protocol. Since the last one year, there have been a couple of proposals on energy models for BLE. However, none of them can model all the operating modes of the protocol. This paper presents a precise energy model of the BLE protocol, that allows the computation of a device's power consumption in all possible operating modes. To the best of our knowledge, our proposed model is not only one of the most accurate ones known so far (because it accounts for all protocol parameters), but it is also the only one that models all the operating modes of BLE. Furthermore, we present a sensitivity analysis of the different parameters on the energy consumption and evaluate the accuracy of the model using both discrete event simulation and actual measurements. Based on this model, guidelines for system designers are presented, that help choosing the right parameters for optimizing the energy consumption for a given application.
\end{abstract}

\fi

\section{Introduction}

Optimizing energy consumption is a crucial design requirement in many wireless sensor networks. Long battery lifetimes are especially important for body-worn medical sensors, mobile phones and interface devices such as a wireless mouse. 
To address this issue, recently an ultra-low-power communication protocol, called Bluetooth Low Energy (BLE) has been proposed \cite{bleSpec}.
BLE-based sensors are able to operate on a coin cell for several months to several years \cite{tian092:12}, depending on its processing and communication demands and the parametrization of the (BLE) protocol. The protocol leaves open a large degree of freedom in terms of choosing the parameter values. Such parametrizations have a significant impact on the power consumption of the device using the protocol. Along with the power consumption, important aspects of communication such as latency and throughput are influenced by this parametrization. For example, the throughput ranges from less than one bit per second (on average) to up to $102 kBit/s$\footnote{A maximum throughput of approximately 40 kBit/s was reported in \cite{mackensen:12} and 54.8 kBits/s was reported in \cite{goemz:12} with a CC2540 BLE device. We could measure $102 kBit/s$ on a CC2540-based BLE112 device.
A derivation in \cite{gomez:11} results in a theoretical maximum possible throughput of 236.7 kBit/s.}, making BLE suitable for a broad range of wireless applications, especially when only small volumes of data have to be transmitted on an average.

As a result, appropriate parameter optimization that takes the application's requirements into account is an important part of the design process. Furthermore, intelligent power management algorithms running on BLE devices can perform online parameter optimizations and therefore need to make use of an energy model.

In this paper, we present a precise and comprehensive energy model for BLE in all its possible modes of operation. Our work integrates existing models for BLE and makes a number of new contributions:
\begin{enumerate}
\item Our model takes into account \emph{all} relevant operating modes and parameters of the protocol, whereas all known models till date only capture a limited set of operating modes. Towards this, we combine, refine and extend a number of recently proposed energy models for BLE.
\item We analyze the variations of parameter values and perform a sensitivity analysis to study their impact on power consumption.
\item We validate the model through detailed experiments, viz., by comparing the results from our proposed model with those obtained from measurements. Such comparisons have not been done for any of the previously proposed energy models, baring one exception in \cite{liu:12_techrep}, where only one mode was studied.
\end{enumerate}


The rest of this paper is organized as follows. In Section \ref{sec:ble}, we give a short introduction to the BLE protocol. Related previous work and our contributions are described in Section \ref{sec:related_work}. We then present our proposed energy model for BLE in Section \ref{sec:bleEmod}. This model is compared against simulations and real world measurements in Section \ref{sec:verification}. In Section \ref{sec:results_and_conclusion}, the model is used to provide guidelines for appropriate parametrization of the protocol in different scenarios. To ease the understanding of the equations we present in this paper, a table of symbols is provided in the appendix.

\section{The BLE protocol}
\label{sec:ble}
\textsc{}In this section we give a short overview of the BLE protocol and describe all aspects of BLE that are important for deriving a power model.
\subsection{BLE versus Bluetooth BR/EDR}
Originally, Bluetooth was developed as a wireless replacement for cables targeting mobile devices. Most of the data transmitted were pictures, music and video files. For these applications, a high speed data link was necessary, which contradicted the goal of extending the lifetime of battery-powered devices. The second type of applications involved connecting computer peripheral devices such as keyboards and mice to a host PC. In this case, the data rate was not a limiting problem, but the battery lifetime had to be extended enormously. To achieve this, the so called \emph{Sniff-Mode} was introduced. For small amounts of data, rather than immediate data transmission, meeting soft deadlines is sufficient. Bluetooth data transmission is organized in time slices and the Sniff-Mode allows data transmission only in a subset of them. Based on the volume of data, the slave sleeps and wakes up periodically at the appropriate time slices. However, in doing so, the master (e.g. a computer) of the Bluetooth connection is always awake, thereby incurring a high energy consumption.

As mobile devices, especially mobile phones, became more powerful, applications using sensor data became more common and all of them used standard Bluetooth even though it was not designed for this purpose, since the master always needs to be awake. In addition, standard coin cell batteries that are used in many setups cannot sustain the high peak currents resulting in Bluetooth BR/EDR. To address these two issues and achieve higher data rates, Bluetooth evolved in two directions: 1) Bluetooth High Speed (HS) which has a higher data rate for optimized data exchange and 2) BLE which targets towards small and low cost sensors powered by coin cell batteries. BLE was developed from its predecessors Wibree and ULP Bluetooth. These were introduced in recognition of the need of an ultra low power wireless data link in addition to the standard Bluetooth BR/EDR.

\subsection{BLE Protocol Fundamentals}
\label{sec:protocol_fundamentals}
\begin{figure}[htb]
\centering
\includegraphics[width = 13cm]{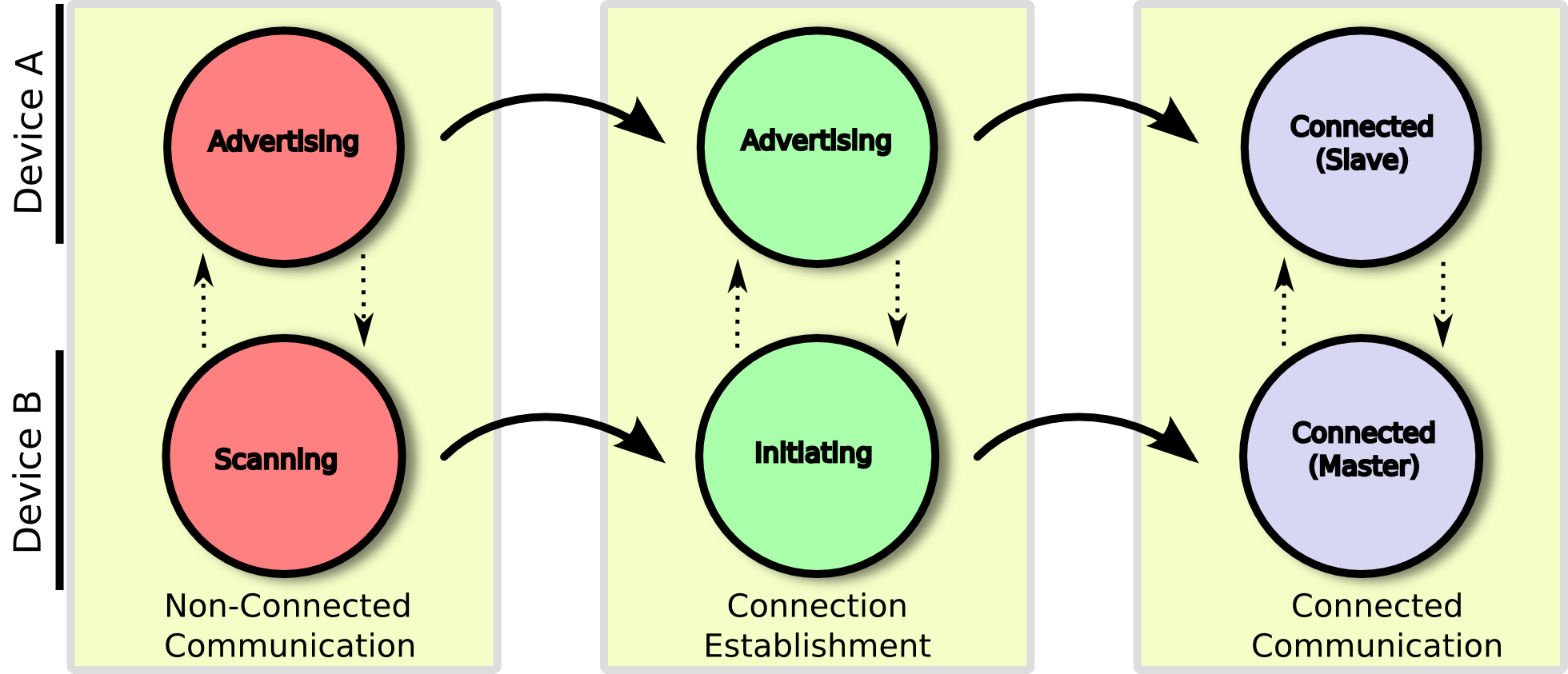}
\caption{Different modes in BLE}
\label{fig:ble_highlevel_schematic} 
\end{figure}
Here, we summarize the main features of the BLE protocol that are relevant for our proposed energy model. More details may be found in the BLE specification \cite{bleSpec}.
In BLE, wireless communication is accomplished on 40 RF-channels on the ISM-band around 2.4 GHz. Three of these channels are reserved for non-connected communication, as described below, whereas the other 37 channels are used in the \emph{connected mode}. On top of this physical layer, the BLE specification describes a protocol that allows devices implementing it to switch to a sleep mode. By allowing a device to wait in such a sleep mode for a large fraction of its operating time, the power consumption is significantly reduced.
Consider two devices A and B that perform a bidirectional wireless communication, as depicted in Figure \ref{fig:ble_highlevel_schematic}.
First, each of them assumes that there is another device in range and thus, both nodes need to discover each other.
One device periodically sends advertising packets, whereas the other device scans for these packets. As soon as the scanner has received at least one advertising packet, it can go to the \emph{initiating} state to establish a connection. In this state, handshaking packets are exchanged, resulting in both devices going to the \emph{connected} state. Once connected, the master controls the timing of the connection and both devices can exchange data in a time-sliced and hence energy-efficient manner.

Even though there are five different states specified by the BLE standard (viz., \emph{Standby}, \emph{Advertising}, \emph{Scanning}, \emph{Initiating} and \emph{Connected}), for the purpose of our model, a more coarse-grained distinction is sufficient. This is because the energy consumption in multiple of these states may be modeled in a similar fashion. As shown in Figure \ref{fig:ble_highlevel_schematic}, we only distinguish between the connected mode, the non-connected mode  and the establishment of a connection. For the simplicity of exposition and without loss of generality, we consider BLE piconets with only two participating devices. However, all of our results can be extended to setups with multiple communicating devices. As a master handles the communication with multiple slaves by exchanging data packets with each device one after another, the energy consumption for each slave can be calculated separately. The only exception is neighbor discovery by advertising and scanning, as multiple advertisers might interfere with each other, leading to packet loss and resulting in a higher energy consumption for neighbor discovery. In addition, some model parameter values might vary for multiple slaves in the connected mode. These issues are not examined in this paper and need to be addressed in future research.

\subsubsection{Non-Connected Communication}
\label{sec:non_con_comm}
Non-connected communication is mainly used for neighbor discovery. Here, one device is in the advertising mode and the other one is in the scanning mode. At the beginning, both nodes are completely unsynchronized. Nevertheless, data exchange is possible by the protocol described below. Advertising and scanning are used both for transmitting information (e.g., related to the services provided by the advertiser) as well as for connection establishment.
\begin{figure}[htb]
\centering
\includegraphics[width = 13cm]{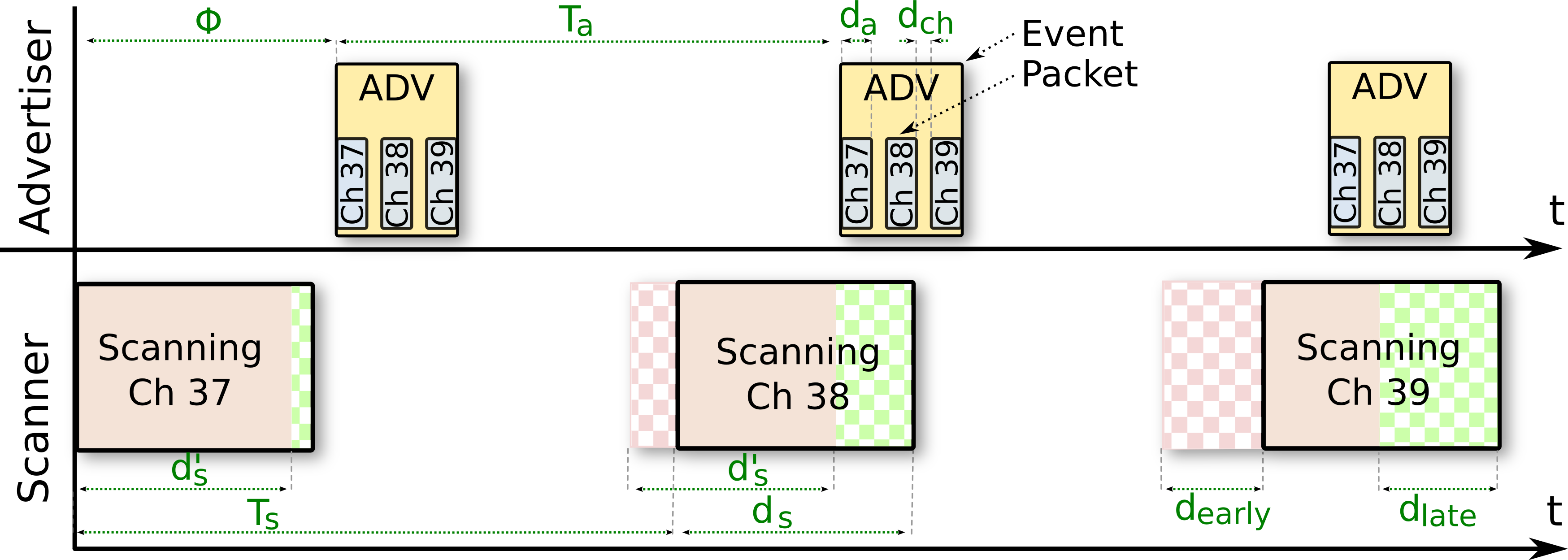}
\caption{Advertising and scanning in BLE.}
\label{fig:advScanning} 
\end{figure}
Advertising and scanning take place as shown in Figure \ref{fig:advScanning}: An advertising node periodically sends out advertising packets. A group of consecutive packets form an \emph{advertising event}. In each of these advertising events, an advertising (ADV) packet is sent to at least one of three dedicated adverting channels. Which of the three channels are to be used is determined by the application.
Advertising events happen regularly with the \emph{advertising interval} $T_a$. 
Independant from this, a scanner periodically switches on its receiver for a duration of $d_s$ time units called the
\emph{scan window}. This is repeated at every interval $T_s$, called the \emph{scan interval}. In the next period, the scanner hops to the next advertising channel and again listens for advertising packets. The specification \cite{bleSpec} requires the scanner to use all three advertising channels. If the scanner receives an advertising packet, it may send a response packet to the advertiser within the same ADV-event. The advertiser expects a response on the same advertising channel $d_{IFS} = 150 \mu s$ after the end of the advertising packet. $d_{IFS}$ is called the \emph{interframe-space}. This enables the following three applications: 
\begin{itemize}
\item \textbf{Neighbor discovery:}
The scanner may respond to an advertising packet with a connection request packet. Both devices will then go into a connected mode, using the procedure described in \mbox{Section \ref{sec:connectionProcedure}}.

\item \textbf{Broadcasting:}
The advertiser may send payload information within its advertising packets. The scanner may receive them in a passive manner without responding.
\item \textbf{Active scanning:}
If the scanner receives an advertising packet, it may send a scan-request-packet $d_{IFS}$ time units after the reception of an advertising packet, requesting more information to be sent by the advertiser another $d_{IFS}$ time units later.
\end{itemize}
The advertising interval is composed of a static interval $T_{a,0}$ and a random part $\rho$, i.e.,
\begin{equation}
T_a = T_{a,0} + \rho,
\end{equation}
where $\rho$ is a random amount of time between $0 ms$ and $10 ms$.
The other timing parameters can be chosen by the host and are required to be within the following time intervals (some additional constraints depending on the mode of operation might apply. We refer to the BLE specification \cite{bleSpec} for details): 
\begin{align}
20 ms & & < & & T_{a,0}& &<& &  10.24 s \nonumber\\
0 ms & & < & &T_s& &<& & 10.24 s \nonumber\\
0 ms & & < & & d_s& &\le& & T_s
\end{align}
The duration of an advertising event can be calculated in a manner similar to that of calculating the duration of a connection event in the connected mode.
The modeling of these events is described in Section \ref{sec:singleEventModel}. The random offset $\rho$ is added to $T_{a,0}$ to avoid the unfavorable case that $T_s = T_a$. In this case, if one advertising event misses a scan event, all subsequent advertising events would miss all future scan events too and therefore the scanner would never discover the advertiser.

As the BLE protocol can be described as a temporal sequence of events, its resulting energy consumption can be modeled by analyzing the energy consumption of its events and their
frequency of occurrence. The energy model proposed in this paper follows this reasoning.

\subsubsection{Connection establishment}
\label{sec:connectionProcedure}

\begin{figure}[htb]
\centering
\includegraphics[width = 13cm]{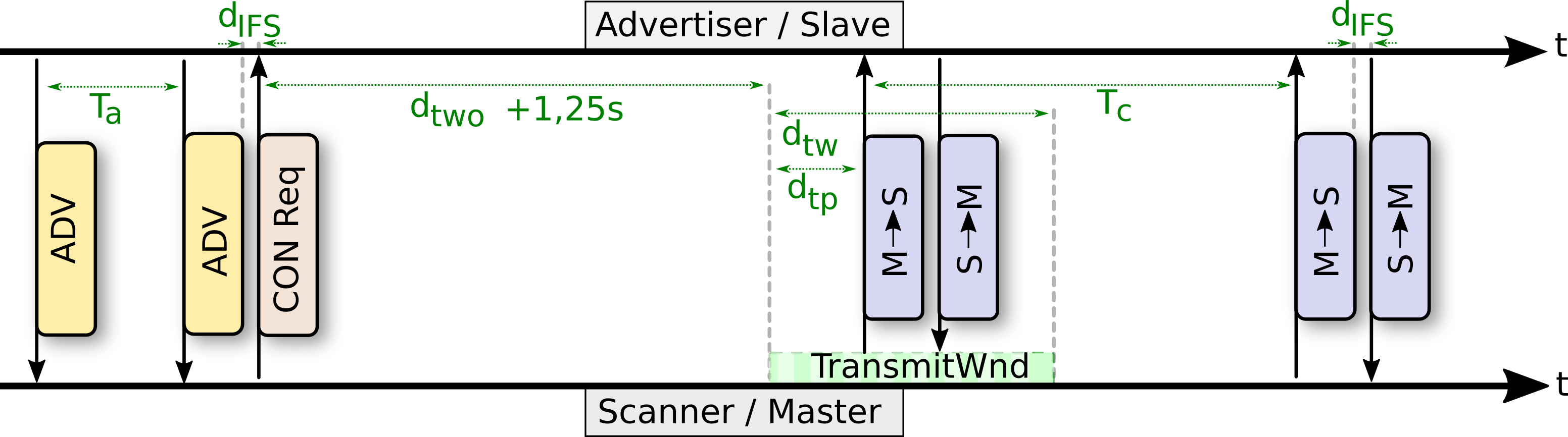}
\caption{Connection establishment packet flow.}
\label{fig:connectionEstablishment} 
\end{figure}

A connection can be established after a device has been detected by advertising/scanning. After having received at least one advertising packet, the scanner sends a connection-request packet $d_{IFS}$ time units later. This packet contains two parameters called \textit{transmitWindowOffset} $d_{two}$ and \textit{transmitWindow} $d_{tw}$ that determine the timing of the connection establishment procedure. In doing so, the scanner is in the initiating state and is therefore called the \emph{initiator}. The durations $d_{tw}$ and $d_{two}$ affect the connection establishment procedure as depicted in \mbox{Figure \ref{fig:connectionEstablishment}}. $1.25ms + d_{two}$ time units after the end of the connection-request packet, the initiator may schedule its first regular connection event, 
defining the anchor point $t_{anchor}$ for all further connection events. The first event may be scheduled at any point in time within the transmit window. Therefore, the advertiser has to listen during the entire transmit window until there is a successful reception. Afterwards, the connection is valid and data can be exchanged as described in Section \ref{sec:connectedMode}. 
As described in the next section, in the connected mode, data is transmitted periodically with the \emph{connection interval} $T_c$.
The procedure described for connection establishment is also used by the master for changing the parameters of an existing connection with some small modifications: In a regular connection event, the master sends the new values for $T_c$ along with other updated parameters to the slave. The transmit window for the first packet being affected by the new parameter values starts $d_{two} + T_{c,o}$ time units after the transmission of the connection update packet has started. $T_{c,o}$ is the connection interval before the parameter update procedure.

\subsubsection{Connected Communication}
\label{sec:connectedMode}
During the connection establishment phase, both devices have agreed on a connection interval $T_c$ that determines the period with which the connection events occur. In addition, there is an anchor point $t_{anchor}$ that both devices have been synchronized on.
As mentioned before, one of the main reasons behind the low power consumption of a BLE device is its ability to sleep for long durations. This is achieved by an appropriate protocol, that is a contract between the \mbox{master \textit{M}}
and the \mbox{slave \textit{S}}. This contract specifies the points in time both devices have to be awake.
At $T_{anchor} + k T_{c}, 0 \le k \le \infty$, both devices have to be awake and \textit{M} will send a packet. \textit{S} will then acknowledge it by sending another packet 150 $\mu s$ later, which might be either an empty response, a payload packet or a
packet performing control functions. If there is more data to be transmitted, more pairs of packets are exchanged in the same manner, separated from each other by $d_{IFS}$ time units. Apart from pre- and postprocessing, the device may sleep during all other times. In addition, there is a parameter called slave latency $N_{sl}$ that increases the sleep duration of \textit{S} in case there is nothing to signal in a connection event.
If \textit{M} and \textit{S} have agreed on a slave latency of $N_{sl}$, the slave might skip $N_{sl}$ connection events of \textit{M} without waking up. The connection interval $T_c$ must range from $7.5 ms$ to $4.0 s$ and $N_{sl}$ may be up to 500 events.
\begin{figure}
\centering
\includegraphics[width = 13cm]{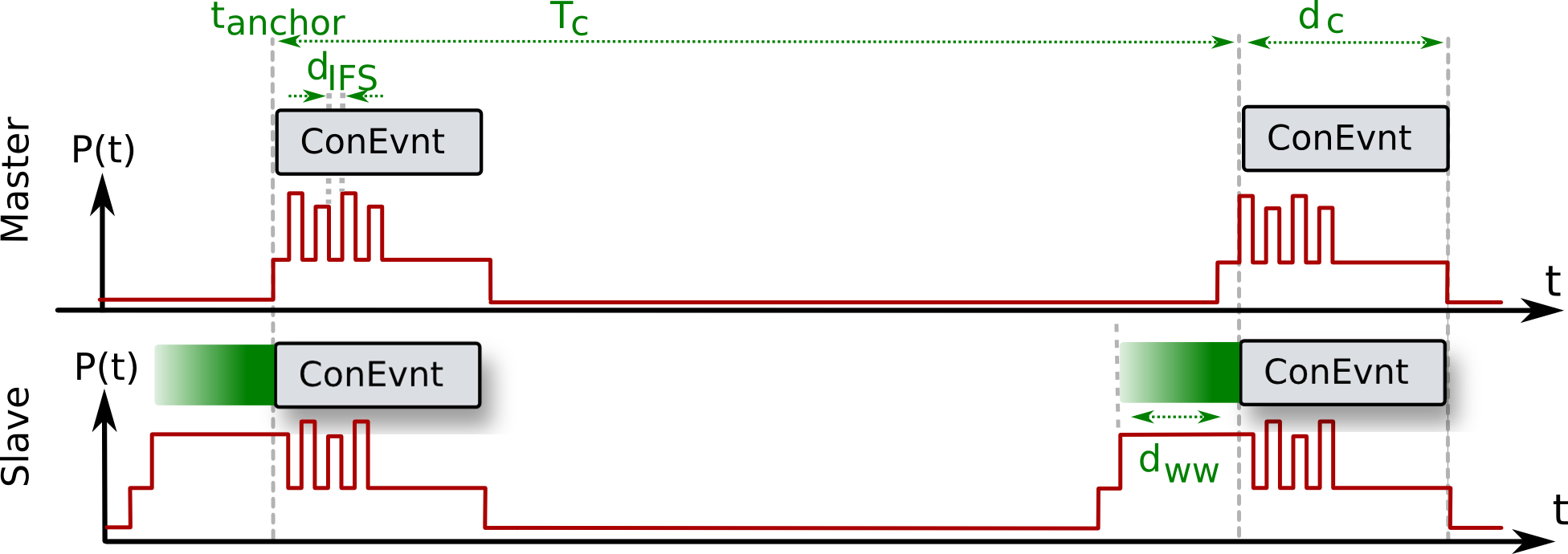}
\caption{Connection events in the connected mode with corresponding power curves P(t)}
\label{fig:connectedMode} 
\end{figure}
Figure \ref{fig:connectedMode} illustrates the concept of connection events. The duration $d_c$ of a connection event can vary from event to event due to varying pre- and postprocessing durations, different number of packets per event and different amounts of payload per packet.
As the slave's sleep clock might drift away from \textit{M}'s sleep clock, \textit{S} has to start listening on the radio channel slightly before the transmission starts. The point in time when \textit{M} ideally starts its transmission, $t_{tr,ma}$, equals to $t_{anchor} + k T_c$. The time interval \textit{S} has to start listening before $t_{tr,ma}$ is referred to as \emph{window widening} $d_{ww}$.

\section{Related work}
\label{sec:related_work}
Recently, issues related to BLE energy consumption have been a topic of active research. 
In \cite{ekstrom:12}, an energy-model for Bluetooth BR/EDR in Sniff Mode was presented, that cannot easily be applied to BLE due to differences between the two protocols.

For BLE, different models have been presented lately. The maximum throughput of BLE was modeled in \cite{gomez:11}, taking into account a given bit error rate.
In \cite{tian092:12}, Texas Instruments provided guidelines on measuring the current consumption of its CC2540 BLE device in the connected mode by measuring the current consumption of the partial events and summing them up. Based on this, TI provided an estimate for the battery lifetime if the protocol parameters do not change. By entering the measured event currents and the durations into an Excel sheet provided by TI, the average current consumption may be calculated. This has been the first step towards an energy model for BLE which only took the connected mode into account. In \cite{siekkinen:12}, an energy model in the connected mode similar to TI's scheme was presented and evaluated for different protocol parameters. A similar event-based model was presented for advertising events in \cite{liu:12_techrep}.
There have also been previous studies on the performance evaluation of BLE, which included aspects like energy consumption, latency, memory requirements of the BLE stack, throughput and maximum piconet size \cite{goemz:12}, \cite{mackensen:12}, \cite{mackensen:12_2}. However, no energy model of the form that we propose in this paper has been presented so far.  In addition, the impact of different settings --- like the transmission power --- on the overal energy consumption has not been evaluated. Furthermore, distortions of the current waveform have not been accounted for in previous models. For example, passive elements in the device's power supply line, such as decoupling capacitors and the shunt resistor for the current measurements smoothen the waveform and modify the duration and amplitude of each part of the current waveform. As stated in \cite{tian092:12}, some parameters are subjected to random variations that influence the power consumption. The impact of these variations on the total power consumption has not been evaluated yet. Instead, they have been assumed to be constant. We have measured the variations of these parameters and incorporated the measurements into our model. Finally, to the best of our knowledge, results from any of the previously proposed models have not been compared to measured power curves, bearing one exception for neighbor discovery presented in \cite{liu:12_techrep}. Consequently, the accuracy of these models was still not clear. Our measurements reveal that 
our model is in close proximity with the measured current consumption. The highest error for the average current consumption per connection interval we measured in our experiments was $6,0\%$.

Besides from connected mode communication, neighbor discovery also contributes to the energy consumption. Energy models for this case are difficult to develop, as neighbor discovery is a stochastic process. Nevertheless, solutions to this problem have been presented:
In the protocol \emph{STEM-B}, neighbor discovery is done in a similar way as in BLE. A model for this was presented in \cite{schurgers:02} and applied to BLE in \cite{liu:12_techrep}, \cite{liu:12_short}  and \cite{liu:12_long}, predicting the energy-consumption of the advertiser in a neighbor discovery process. However, it has been stated in \cite{liu:12_long} that this model is only valid for a certain range of possible parameter values ($T_a < d_s$ or $T_s = d_s$ as defined in Section \ref{sec:ble}). In \cite{liu:12_techrep}, measurements comparing modeled and measurement energy consumption have been performed for the neighbor discovery model proposed in that paper. For modeling the energy consumed during neighbor discovery, an energy model consisting of partial events for advertising, similar to the one in \cite{tian092:12} has been used.\\
\newline
\noindent \textbf{Our contributions:}
In this paper, we propose a comprehensive energy model of the BLE protocol, modeling the energy consumption for all operation modes and for all possible parameter valuations. Towards this, we integrate a number of previously known models and at the same time refine and extend them in order to obtain a complete model that accounts for all possible modes of communication with all possible parameters. In particular, our contributions are as follows.
\begin{enumerate}
\item To the best of our knowledge, we present the first model that includes i) all possible modes of operation, ii) all relevant parameters, and iii) all possible parameter valuations of the BLE protocol. For example, in the connected mode, our model takes into account the distortion of the current curve caused by resistive and capacitive elements present in the power supply line. Instead of relying on theoretical data from the BLE specification \cite{bleSpec} and values from device data sheets only, our model includes measured durations and current magnitudes, leading to more accurate predictions on state durations compared to previous models. For example, in our model the duration of the transmission state is modeled as a constant part and a part depending on the number of bytes sent rather than a byte-dependent duration only. In addition, we present the first model for scan events, for connection setup and for updating the connection parameters. We are also the first to present an algorithm that estimates the discovery latency and energy consumption for neighbor discovery for all possible parameter values, including $T_a > d_s$. Analytical solutions exist only for $T_a < d_s$ \cite{liu:12_techrep}. Based on the proposed model, we present guidelines to system designers for determining protocol parameter values that optimize the power consumption for different application scenarios.
\item Using multiple measurements, we identified the degree of variation in the valuation of the different parameters and the sensitivity of these parameters on the total energy consumption. In many cases, it is reasonable to use mean values of varying parameters, but sometimes this is insufficient. For example, when estimates on maximum current consumptions over small time intervals are necessary, identifying a range over which the parameter varies is required. In a sensitivity analysis, we investigated the impact of these variations on the overall power consumption. In addition, we analyzed the impact of different transmission-power-settings on the energy consumption.
\item To the best of our knowledge, none of the known energy models for BLE have been compared to measured data yet, bearing one exception presented in \cite{liu:12_techrep}, where the energy consumption of an advertiser for neighbor discovery was modeled and compared to measured data.
We have validated the results from our proposed model with real experimental results. Therefore, we performed comparative measurements for the modeled neighbor discovery latency to prove that our results are in very close proximity to the measured data.
In measurements we performed for the connected mode, the modeled mean current consumption within a connection interval differed from the measured results by between $1.17 \%$ and $6.0 \%$.  
\end{enumerate}

\section{BLE Power Model}
\label{sec:bleEmod}
In this section, we present a precise energy model for BLE that is capable of predicting the energy-consumption 
of the protocol in all modes of operation.
\subsection{Overview}
\begin{figure}[htb]
\centering
\includegraphics[width = 13.0cm]{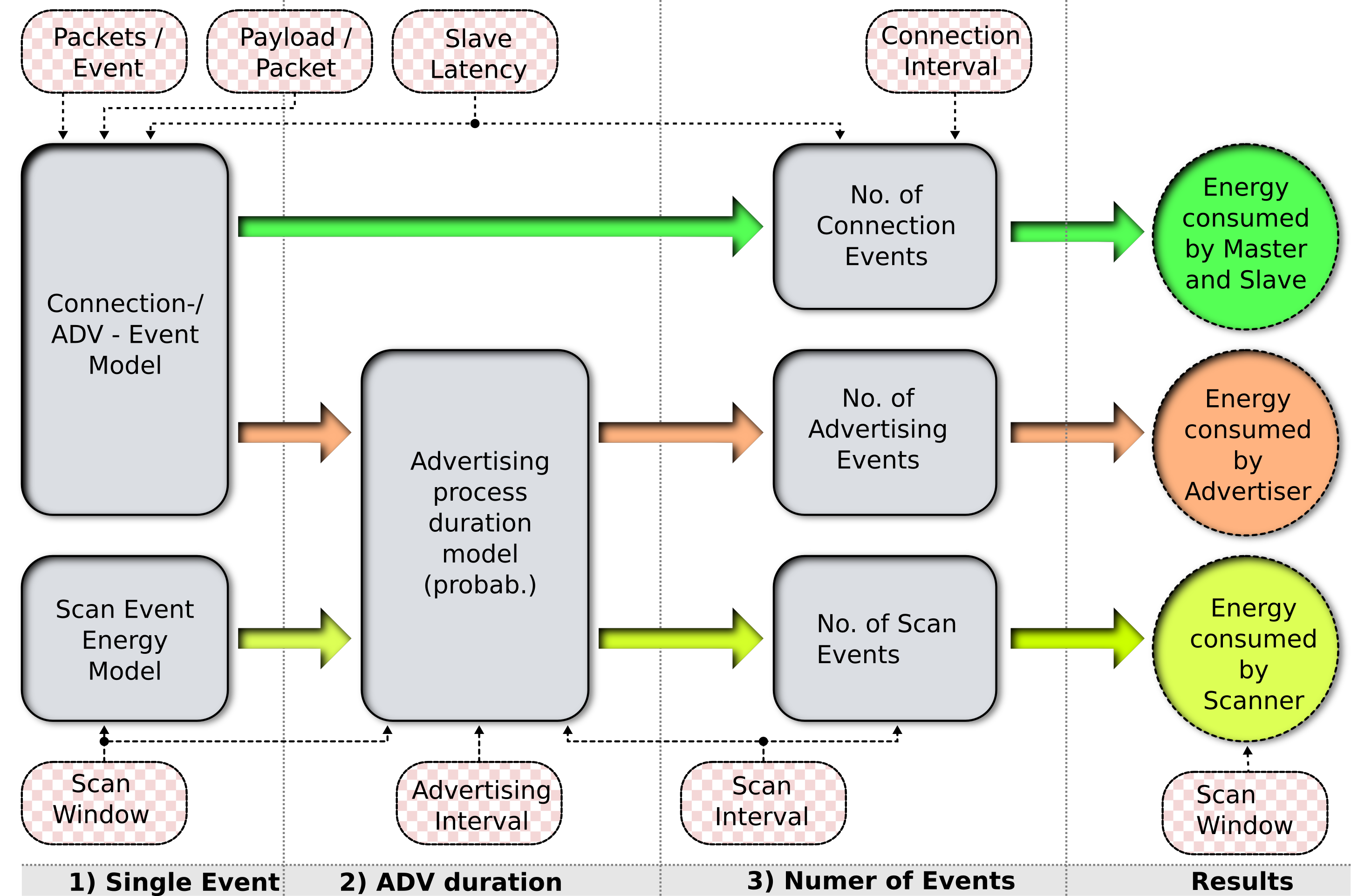}
\caption{High-level diagram of the BLE energy model. The hatched boxes denote BLE protocol parameters used by the model.}
\label{fig:model_highlevel} 
\end{figure}
Figure \ref{fig:model_highlevel} gives an overview on our approach.
The BLE protocol can be modeled as a temporal sequence of single events that happen periodically. Energy models of these events depend on their type (connection- , advertising- or scan-event), as can be seen on the left in Figure \ref{fig:model_highlevel}. The hatched boxes in the figure depict protocol parameters that are fed into the model. 

In Stage 1) of Figure \ref{fig:model_highlevel}, the energy of all single events that occur is modeled. In Stage 2), the average duration of an advertising/scanning-process $\overline{d_{adv}}$ is modeled for the advertising/scanning mode. This duration is of stochastic nature. It determines how often an advertising event or scan event has to be repeated before the data (either payload or handshaking for the initialization of a connection establishment process) has successfully been received at the receiving node. In the connected mode, this time is deterministic and is defined by the connection interval $T_c$ and the slave latency parameter $N_{sl}$. In the third stage, the number of identical events that occur within a certain amount of time is calculated. Next, the energy consumed per event is multiplied by the number of times it occurs and the results for different events are then summed up. In addition, the time the device sleeps between these events and its corresponding sleep energy consumption is calculated. By combining the energy of all events with the energy spent during sleeping, the expected energy for transmitting a given payload for a given set of parameters can be calculated. 
In this section, we describe every block in Figure \ref{fig:model_highlevel}.
In addition, we describe an energy model for the connection establishment procedure, which is not shown in the figure. This procedure occurs whenever the scanner/initiator has discovered its remote device and establishes a connection or connection parameters are to be updated in an existing connection.

\subsection{Single event model}
\label{sec:singleEventModel}

\subsubsection{Connection/Advertising Events}
\label{sec:conAdvEvent}
In this section, we present a model for connection events. As already mentioned, 
the energy consumption of an advertising event can be modeled similarly as both events draw similar currents. Therefore, the model for connection events of a master, as described in this section, can be used to model advertising events too.
A previous energy model for connection events in BLE has been presented in \cite{siekkinen:12}. In our model we extended this by taking into account window widening, non-ideal durations, a communication preamble and correction terms to account for distortions of the current curve caused by resistive and capacitive elements in the power supply line. As the model is based on measured data, these effects must be taken into account for making accurate predictions.
Finally, we subdivide the connection event in finer-grained phases to model the energy consumption as precisely as possible.

Like the CC2540 BLE module \cite{an097}, the BLE device we analyzed (Bluegiga BLE112, based on Texas Instruments CC2540) makes use of an internal low-dropout voltage regulator (LDO), which keeps the current consumption independent from the supply voltage. For this reason, we present all parameter values of our model in terms of electric current $I$ in amperes and electric charge $Q$ in coulombs. From these values, the power- and energy-consumption can be easily obtained for a given supply voltage.
The description in this section is made for a BLE slave. For a master, the \emph{rx}- and \emph{tx}-phases described below are interchanged and no window-widening occurs. Therefore, the model described can be used for a BLE master too.

\begin{figure}
\centering
\includegraphics[width = 13.0cm]{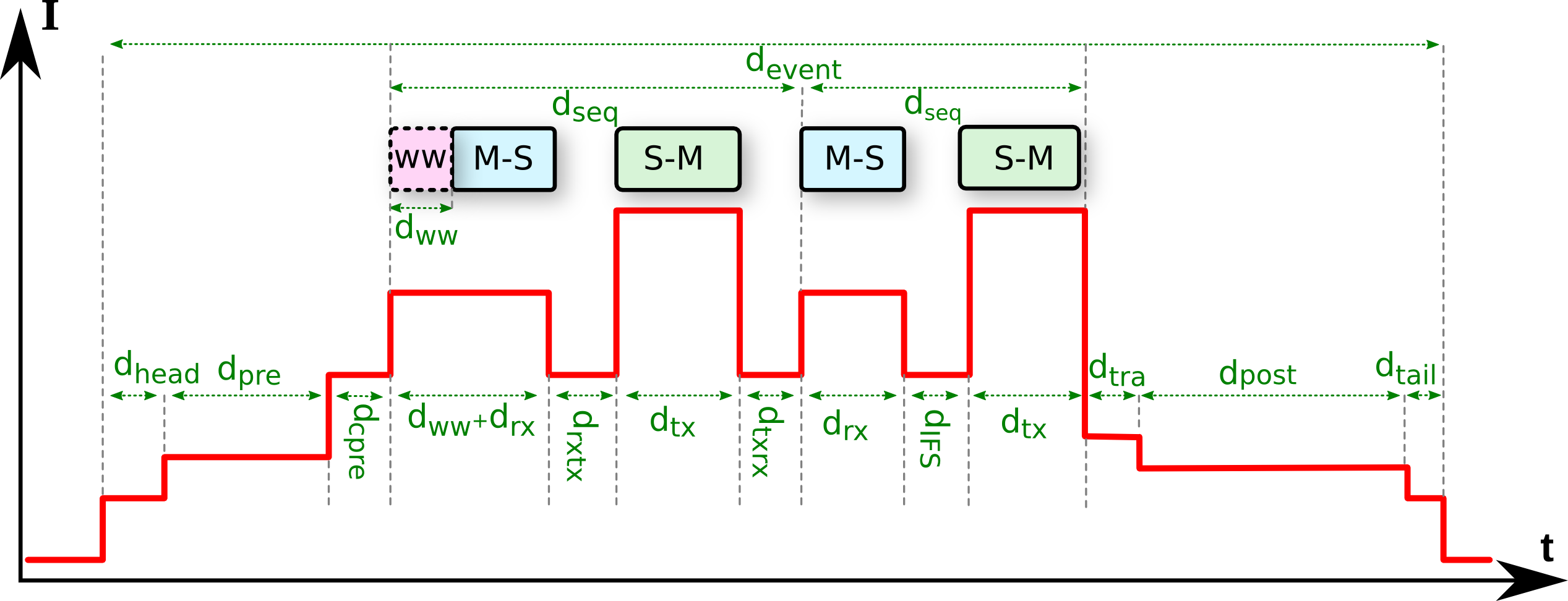}
\caption{Model for a BLE connection event of a slave device.}
\label{fig:packet}
\end{figure}

Within a connection event, ten distinguishable phases with nearly constant current consumptions can be identified. As shown in Figure \ref{fig:packet}, these phases are as described below.

\begin{enumerate}
\item \textbf{Header ($head$):}
At the beginning of a connection event, the device wakes up. Thereby, a current $I_{head}$ is consumed for the duration of $d_{head}$.
\item \textbf{Preprocessing ($pre$):}
Prior to the actual communication, the device wakes up and prepares the functions related to the BLE standard, which are used during the communication, such as the logical link control and adaptation protocol (L2CAP), generic access profiles (GAP), generic attribute profile (GATT), security manager (SM), etc. While the average current is nearly constant over time during this phase, its duration might vary depending on the BLE device, the BLE stack and the BLE functionality that is used (.e.g., sending attributes with or without confirmation). In addition, random variations occur, as can be seen in the tables of Appendix \ref{sec:appendixParamsConnected}.
\item \textbf{Communication preamble ($cpre$):}
The wireless communication is initiated in this phase. It is inherent to the hardware and its current and duration may change with different BLE device models.
\item \textbf{Window widening ($ww$):}
As already mentioned, the slave has to start listening for a certain amount of time before the master starts sending its first packet to compensate for the time drift that might be generated due to clock skews. This additional time is called \emph{window widening} $d_{ww}$. It depends on the nominal sleep clock accuracy (SCA) of each transceiver, the connection interval $T_{c}$ and the average slave latency $\overline{N_{sl}}$ as shown in Equation \refeq{eq:windowWidening}. The sleep clock is the clock that determines the points in time for waking up a sleeping device. The current magnitude of this phase is the same as the magnitude of the \emph{rx}-phase.
The duration $d_{ww}$ can be calculated according to Equation \refeq{eq:windowWidening} \cite{bleSpec}, assuming that the average clock skew is equal to zero (positive and negative skew compensate for each other).
\begin{align}
\label{eq:windowWidening}
d_{ww} &= \frac{(SCA_{Ma}+SCA_{Sl})\cdot T_{c} \cdot \overline{N_{sl}}}{10^6}\text{, } & I_{ww} = I_{rx}
\end{align}
\item \textbf{Reception ($rx$):}
\label{it:reception}
The over-the-air-bitrate of BLE is specified to be $1\mbox{ }MBits/s$, therefore one bit is transmitted within $1 \mu s$. Consequently, the \emph{rx}-phase should ideally take $N_{rx} \cdot 8 \mu s$, with $N_{rx}$ being the number of bytes received.
Because the RF circuitry needs some time to initialize, the duration of the \emph{rx}-phase is slightly longer than its ideal value. To account for this, a correction-offset $d_{prerx}$ is added to $d_{rx}$ as in the equation below.
\begin{equation}
d_{rx} = N_{rx} \cdot 8 \mu s + d_{prerx}. 
\end{equation}
The current $I_{rx}$ is nearly constant; some devices such as the CC2540 \cite{CC2540_datasheet} have multiple reception-gain settings that cause different current draws.
\item \textbf{Interframe-Space ($rxtx$, $txrx$):}
After the \emph{rx}-phase and before the slave starts sending a packet to the master, there is a phase for switching from reception to transmission and vice-versa. Its duration is slightly shorter than the over-the-air gap between two packets $d_{IFS}$. 
\item \textbf{Transmission($tx$):}
In this phase, the slave transmits data to the master. Its duration can be modeled in a manner similar to the \emph{rx}-phase using Equation \refeq{eq:tx}.
\begin{align}
\label{eq:tx}
d_{tx}&= d_{pretx}+N_{tx}\cdot 8\mu s
\end{align}
$N_{tx}$ is the number of bytes transmitted. The current consumption $I_{tx}$ depends on the \emph{tx}-power that is used. For a BLE112-device, there are 15 different \emph{tx}-power-settings. In addition, $I_{tx}$ slightly varies with the channel a packet is sent on. As the channel is determined by a pseudo-random hopping sequence, these variations can be assumed to occur randomly with a given standard deviation.
The BLE protocol allows the transmission of multiple pairs of packets in a single connection event. After another \emph{txrx}-phase, the \emph{rx}, \emph{rxtx} and \emph{tx}-phases might be repeated to account for multiple pairs of packets.
\item \textbf{Tx transient ($tra$):}
After the data transmission has ended, the current decreases from $I_{tx}$ to the postprocessing-current $I_{post}$. Whereas the current consumption drops with a RC curve, an effective constant current $I_{tra}$ and an effective duration $d_{tra}$ can be chosen appropriately to take the charge consumed by this into account.  
\item \textbf{Postprocessing ($post$):}
After the communication has ended, the device may execute additional tasks such as wired communication to a host processor or data buffering for the next transmission. Therefore, the duration of this phase is strongly dependent on the BLE device and its software. In our experiments, we measured the post-processing duration caused by a firmware created with Bluegiga's BGScript \cite{ble112} without accounting for additional tasks. Both on BLE112- and CC2540-devices, $d_{post}$ is subjected to strong random variations \cite{tian092:12}. When executing one BLE functionality (e.g., sending a packet), we measured varying postprocessing times as presented in the tables in Appendix \ref{sec:appendixParamsConnected}.
Correlations between $d_{pre}$ and $d_{post}$ might exist. In \cite{tian092:12} it is stated that for a CC2540-device, the sum of $d_{pre}$ and $d_{post}$ is always the same. In our measurements for a BLE112-device, we could not reproduce this behavior\footnote{For example, the sum of $d_{pre}$ and $d_{post}$ varied for different numbers of packets per event.}. For our model, we assume that there is no correlation between $d_{pre}$ and $d_{post}$. If precise predictions on short-time currents need to be done, one should analyze the correlations of these durations for the device used.
\item \textbf{Tail ($tail$):}
After the BLE device has completed the tasks related to the post-processing, it goes to a sleep mode to reduce the energy consumption. This phase in the model accounts for the current consumed for initiating the low power mode.
\end{enumerate}
The timing for these phases differ for the over-the-air traffic and the current-curve measured. The values needed for an energy-model consider the timing of the current draw.

For each phase $ph$, the charge consumed can be calculated by $Q_{ph} = d_{ph} I_{ph}$.
With the phases described, the charge consumed for a connection event, $Q_{cE}$, can be computed as in Equation \refeq{eq:qConEv}.
\begin{align}
\label{eq:qConEv}
Q_{cE} &= Q_{head} + Q_{pre} + Q_{cpre} + Q_{ww} + Q_{t} + Q_{tra} + Q_{post} + Q_{tail}
\end{align}
$Q_{t}$ accounts for the actual communication taking place. For a communication with $N_{seq}$ pairs of packets, it is
\begin{align}
Q_{t} = \sum_{i=1}^{N_{seq}}(Q_{rx}(i) + Q_{tx}(i) + Q_{rxtx} + Q_{txrx} + Q_{to}) - Q_{txrx}
\end{align}
$Q_{to}$  is an offset to account for distortions in the current curve. Due to the distortion of the current curve caused by resistive and capacitive elements in the power supply line, the shape of the phases of the current curve are not perfectly rectangular. This can be compensated with effective parameter values for event-phases with constant length. In contrast, for the communication sequence, an offset term needs to be added as the varying durations make it impossible to find effective values accounting for offsets that occur.

The duration of a connection event can be calculated similar to Equation \refeq{eq:qConEv} by adding the partial durations of all phases.

\subsubsection{Scan Event}
\label{sec:scanEvent}
The current waveform of a scan event depends on the mode the device is in. Possible modes are described in Section \ref{sec:non_con_comm}.
We describe the energy consumption for an event with active scanning, which is the most complex waveform, and simplify it for events with no reception of data and for events which receive connection-request packets.
\begin{figure}[htb]
\centering
\includegraphics[width = 13cm]{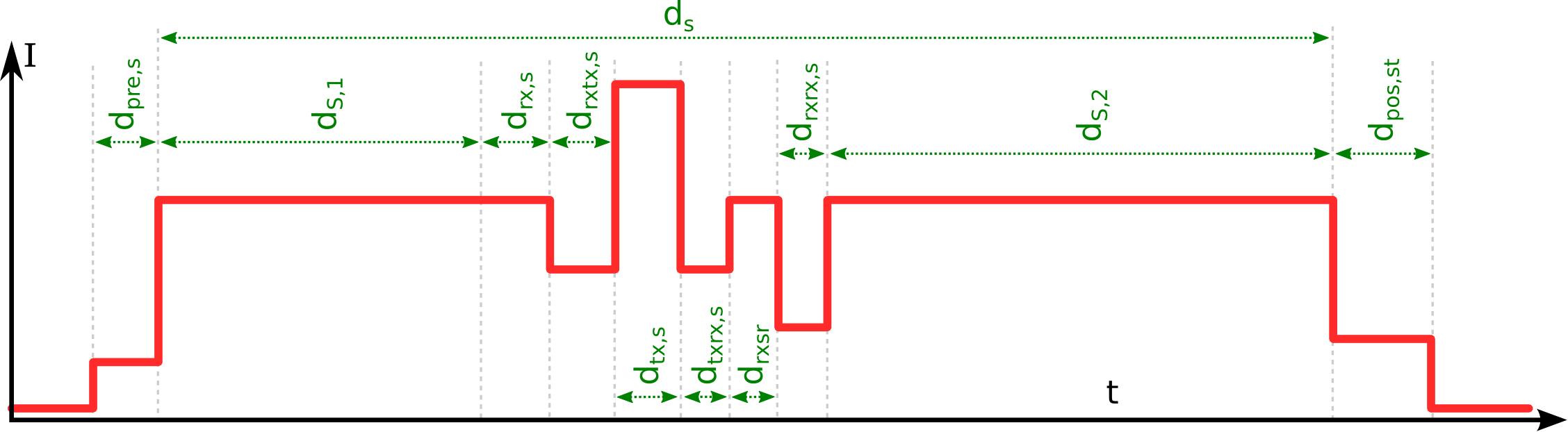}
\caption{Current waveform of a scan event when receiving and serving a scan request packet (active scanning)}.
\label{fig:activeScanning} 
\end{figure}
In the equations in this section, paramters denoted with an \emph{s} indicate \emph{scanning}, as the parameter values differ from the values in the connected mode.
\mbox{Figure \ref{fig:activeScanning}} shows the current-waveform for active scanning, assuming that exactly one advertising event is received during one scan event. As for the connection event, the device wakes up, drawing a current for the duration of $d_{pre,s}$\footnote{We denote the phase at the beginning of a scan event as the preprocessing phase and the phase at the end of a scan event as the postprocessing phase. This does not mean that the device necessarily does data processing in these phases.}.
Afterwards, it scans for the scanning time $d_{S,1}$ until an incoming advertising packet is received, consuming a high current $I_{scan}$ = $I_{rx,s}$ due to the permanent usage of the receiver circuit. During the reception of the packet, the current consumption remains the same.
The device then switches from receiving to transmission, which takes $d_{rxtx,s}$ amounts of time. Next, it sends a scan request packet lasting for $d_{tx,s} = N_{tx} \cdot 8 \mu s + d_{pretx,s}$ time units. Here, $N_{tx}$ is the number of bytes sent and $d_{pretx,s}$ is a correction offset accounting for charging and discharging capacitances. Afterwards, the devices switches from \emph{tx} to \emph{rx}, taking the time $d_{txrx,s}$ and subsequently, a scan response packet from the remote device is received. This takes $d_{rxsr} = N_{rx} \cdot 1 \mu s + d_{prerx,s}$ time units. Before the device continues scanning, there is a phase with constant current for $d_{rxrx,s}$ units of time. When the time $d_{S,2}$ has expired, the device stops scanning and goes to sleep, consuming a smaller current for the time $d_{post,s}$ before the sleep phase begins.
The scan durations $d_{S,1}$ and $d_{S,2}$ depend on the time an advertising packet is received and hence are random. Fortunately, for calculating the energy consumption, only the sum of $d_{S,1}$ and $d_{S,2}$ is of interest, which is given by the scan window $d_s$. It is approximately $d_{scan} = d_{s} - d_{rxtx,s} - d_{tx,s} - d_{txrx,s} - d_{rxsr} - d_{rxrx,s}$.
Therefore, the charge consumed by a scan event with active scanning is:
\begin{align}
\label{eq:qsev}
Q_{sEv} = & d_{pre,s} + d_{scan} I_{rx,s} + d_{rxtx,s} I_{rxtx,s} + d_{tx,s} I_{tx,s} + d_{txrx,s} I_{txrx,s} + d_{rxsr} I_{rx,s} \nonumber \\
		  + & d_{rxrx} I_{rxrx} + d_{post,s} I_{post,s}+ Q_{crx,s} + Q_{ctx,s}
\end{align}
$Q_{crx,s}$ and $Q_{ctx,s}$ are correction offsets to compensate for non-rectangular shapes in the reception and transmission phases.
Parameter values for Equation \refeq{eq:qsev} for BLE112-devices are presented in Appendix \ref{sec:appendixParamsConnected}.
In many cases, this equation can be simplified. For sending a connection request packet, the waveform begins similar to what is shown in Figure \ref{fig:activeScanning} but instead of a scan-response packet, a connection request packet is sent. In this case, the \emph{tx}-phase is followed by the postprocessing phase without further sections in between. Our experiments showed that the pre- and post-processing phases for scanning with a connection request last longer than for active scanning on a BLE112-module. This behavior depends on the device and its BLE stack. For the sake of simplicity of explanation, we nevertheless assume these durations to be constant in the values presented in this paper, as this could easily be accounted for by adjusting the values to the given case. Because of the reduced number of phases in the case a connection is initiated, $d_{rxtx,s}$, $d_{rx,s}$, $Q_{crx,s}$  and $d_{rxrx,s}$ can be set to zero and $d_{scan}$ is shortened to the time between the beginning of the idle scanning and the time the advertising packet has been received completely.
If no advertising packet is received or only passive scanning is used, there is only idle-scanning (or the reception of advertising-packets that consumes the same energy as scanning) and the charge consumed can be computed in a simplified way:
\begin{equation}
Q_{sEv,Idle} = d_{pre,s,i} I_{pre,s,i}+ d_{post,s,i} I_{post,s,i }+ d_s I_{rx}
\end{equation} 
If $d_s$ is long compared to $d_{pre} + d_{post}$, pre- and postprocessing can be neglected and the formula above can be simplified further.
As pre- and postprocessing-durations and their effective currents differ significantly between active/idle scanning and scanning with establishing a connection, they are denoted with an index \emph{i} for \emph{idle}\footnote{For the BLE112-stack, our measurements had the following limitations: 1) for scanning with connection-requests, only continuous scanning  ($d_s = T_s$) was possible. 2) the time the device was actually scanning was slightly below $d_{s}$. For scan windows between 12,5 ms and 250 ms, in 1725 analyzed events it was in average approx. 1,85 ms shorter.}.
In many cases, it is hard to predict whether a scan event will receive an advertising packet or not and the point in time the reception takes place. For connection establishment, a solution for this problem is described in Section \ref{sec:advScanningModel}. For active scanning, discrete event simulations can be combined with the model described to get an estimate for the energy consumption.

Another special case is continuous scanning ( $d_s = T_s$): In this case, a scan event does not end with post processing, but the next \emph{rx,S}-phase starts after $d_s$ has passed, having a short phase for channel changing that consumes the charge $d_{chch} \cdot I_{chch}$ in between.

\subsubsection{Connection Procedure}
\label{sec:conProc}
In this section, we propose a model for the energy spent on establishing a connection and, as both procedures are similar, for updating the communication parameters of an existing connection. These procedures are described in Section \ref{sec:conProc}.
To establish a connection, the master sends a connection request packet first, with the energy consumption associated with this event being $Q_{ev,cR,Ma}$. For updating the parameter values of an existing connection, the master consumes $Q_{ev,cU,Ma}$.
Subsequently, there is no communication for i) $d_{sl,cR} = d_{two} + 1.25 ms$ after the end of the connection request packet in the case of a connection establishment, or ii) $d_{sl,cU} = d_{two}$ after the end of the old connection interval in the case of a connection parameter update.
After that, the transmit window begins, as shown in Figure \ref{fig:connectionEstablishment}. Within the size of the transmit-window $d_{tw}$, the master may schedule its first packet, thereby defining the anchor-Point $t_A$ for future connection events.

With some constraints, the master is free to choose the transmit window offset $d_{two}$ between $1.25m$s and $T_{c,n}$, with $T_{c,n}$ being the future connection interval. 
For the transmit window size $d_{tw}$, the master can choose any value\footnote{As integers are used for all parameters in BLE, the values of $d_{tw}$ and $d_{two}$ are quantized and must be a multiple of $1.25 ms$.} in the following range:
\begin{equation}
1.25ms < d_{tw} < min(10ms, T_{c,n} - 1.25 ms)
\end{equation}
With $d_{p} < d_{tw}$ being the duration from the beginning of the transmit window to the time the first packet sent by the master, the charge consumed by the master for establishing a connection can be modeled as:
\begin{equation}
\label{eq:conReqMa}
Q_{cE,Ma} = Q_{ev,cR,Ma} + (1.25ms + d_{two} + d_{p}) I_{sl}
\end{equation}
For a connection update, it is:
\begin{equation}
\label{eq:conUpMa}
Q_{cU,Ma} = Q_{ev,cU,Ma} + (T_{c,o} + d_{two} + d_{p} - d_{ev,cU,Ma}) I_{sl}
\end{equation}
$I_{sl}$ is the sleep current of the BLE device.
$T_{c,o}$ is the connection interval before the connection parameter update took place. $Q_{ev,cR,Ma}$ and $Q_{Ev,cU,Ma}$ are the charges consumed by the events the connection request/update packets are sent within. For an event with connection request packet, these values can be modeled as described in Section \ref{sec:conAdvEvent} as an advertising event with 37 bytes sent by the advertiser and with a response of 44 bytes length sent by the scanner(i.e., connection-request packet). Connection update requests can be modeled as a connection event with a 22-byte packet sent by the master and packets received from the slave depending on the payload the slave has to send.

The master's energy-consumption mainly stems from sending the connection request/update packet to the slave. Opposed to that, the energy consumption of the slave is dominated by the current of its receiver during idle-listening in the transmit window and during the reception of the packet. It can be modeled by the following equations:
\begin{align}
\label{eq:conReqSl}
Q_{cE, S} &= Q_{ev,cR,Sl} + (d_{two} + 1.25ms - d_{ww,cE}) I_{sl} + (d_{p} + d_{ww,cE}) I_{rx}
\end{align}
and
\begin{align}
\label{eq:conUpSl}
Q_{cU, S} =  Q_{ev,cU,Sl} + (T_{c,o} + d_{two} -  d_{ww,Cu} - d_{ev,cU,Sl}) I_{sl} + (d_{p} + d_{ww,cU}) I_{rx}
\end{align}
$Q_{ev,cR,Sl}$/$d_{ev,cR,Sl}$ and $Q_{ev,cU,Sl}$/$d_{ev,cU,Sl}$ are the charges/durations consumed by the events where the connection request is sent/connection update-packet is received within.
As $Q_{ev,cR,Sl}$ is a scan event, it can be modeled according to Section \ref{sec:scanEvent}, whereas $Q_{ev,cU,Sl}$ is a connection event and therefore can be modeled according to Section \ref{sec:conAdvEvent}.
$I_{rx}$ is the current consumption when the receiver listens to the radio channel. The first connection event after this procedure has to be modeled without window widening as the equations above already account for it.
Due to possible clock skew on both devices, window widening occurs that broadens the transmit window and therefore the duration the current $I_{rx}$ is drawn. $d_{ww,cE}$ and $d_{ww,cU}$ contain the times the transmit window is broadened by. With $SCA_{Ma}$ and $SCA_{Sl}$ being the clock accuracies of master and slave in parts per million, the window-widenings for connection establishment $d_{ww,cE}$ and for connection parameter updates $d_{ww,cU}$ are:
\begin{equation}
\label{eq:wwConEq}
d_{ww,cE} = \frac{SCA_{Ma} + SCA_{Sl}}{1000000} (1.25ms + d_{two})
\end{equation}
\begin{equation}
\label{eq:wwConUp}
d_{ww,cU} = \frac{SCA_{Ma} + SCA_{Sl}}{1000000} (T_{c,o}+ d_{two})
\end{equation}
In Equations \refeq{eq:conReqMa} and \refeq{eq:conUpMa}, we neglected the fact that the sleep duration is slightly shorter than what has been modeled, as the first event after the connection establishment or update sequence overlaps with the sleep duration (e.g. header, preprocessing, communication preamble). However, due to the small sleep current of BLE devices, the impact of this is low.
The values of $d_{tw}$, $d_{p}$, $d_{two}$ are chosen freely by the BLE stack and are therefore unknown, making the equations above hard to evaluate. However, a worst case-model can be defined.
Both for the master and the slave, the maximum energy is consumed if both parameters have their maximum values, viz. $d_{p} = min(10ms,T_{c,n} - 1.25ms)$ and $d_{two} = T_{c,n}$. 
Moreover, even if not explicitly defined by the BLE specification, some assumptions on the values of these parameters can be made for real-world devices, that lead to an estimation of the average energy consumed for establishing a connection or updating the connection parameters. Both parameters are used to allow the master to schedule a new anchor point $t_a$. At this point in time, the first connection event of the new/updated connection starts. Using $d_{two}$, the first connection event can be scheduled if it is already known at the time the master sends the connection request/update packet to the slave. In addition, $d_{tw}$ leaves open a time interval for the master to schedule the anchor point without determining it when sending the connection request or update packet.
As there is no reason for the master for not being able to schedule $t_{Anchor}$ already when sending the connection parameters to the master, we assume that it will choose a small values of $d_{tw}$. We further assume that every point in time within the transmit window is taken with the same likelihood. Hence, on an average the master will schedule $t_A$ at the middle of this interval, i.e. $d_{p} = \frac{d_{tw}}{2}$.
If the master choses small transmit windows $t_{tw}$, there is a clear benefit for its energy consumption and as a consequence, a good BLE stack is likely to do so. However, making reasonable assumptions for the value of $d_{two}$ is more difficult. $d_{two}$ only influences the charge spent by sleeping and by window widening. The energy consumption related to this is expected to be small against the energy consumed for idle-listening within $d_{tw}$ and for receiving a packet. From an energy-perspective, the BLE stack can chose an arbitrary value. For energy modeling, $d_{two}$ can be assumed to have its maximum value $T_{c,n}$ if the actual value of the device is unknown.

Whereas a coarse estimation of the energy consumed by the procedure described can be made using these assumptions, for more precise energy estimations, the parameters actually chosen by the BLE stack considered need to be analyzed. We observed the values used by the stack of a CC2540-based BLE112-device manufactured by Bluegiga, by using a Texas Instrument's packet sniffer software \cite{tiSniffer:10}. A passive BLE device was used to wiretap the BLE traffic between two different communicating devices.
We developed a parser for the logs generated by the SmartRF Packet Sniffer software \cite{tiSniffer:10}, allowing for the import of the sniffed data into MATLAB.

By manually experimenting different parameter values, we found that $d_{tw} = 3ms$ is constant for all parameter values in all situations. We further analyzed 26 connection procedures, where $d_p$ had an average value of $1.43 ms$ which is close to $\frac{d_{tw}}{2}$, as we assumed. $d_{two}$ may vary for different values of $T_C$  when establishing a new connection, while being constantly equal to zero for connection updates. We developed a software that repeatedly connects and disconnects two BLE nodes. The packet-sniffer-logs were analyzed to obtain $d_{two}$ for different values of $T_c$. An analysis of about 12,000 connection procedures revealed the following result for a BLE112-device:
$d_{two}$ grows with increasing values of $T_c$, but there are some random variations for the same value of $T_c$. 
For $T_c > 12.5ms$, an estimation for $d_{two}$ is:
\begin{equation}
\label{eq:ble112two}
 d_{two} = T_c - 6.454 ms
\end{equation}
The maximum deviation that occurred in all the measurements was $6.046 ms$ ($0.87 \%$ of the corresponding measured value), the mean square error was $8.756 \mu s$ and the standard deviation $\sigma$ was $2.959 ms$. These results show that Equation \refeq{eq:ble112two} is a good estimation of $d_{two}$.
For $7.25ms < T_c < 12.5 ms$, a good estimation for $d_{two}$ is:
\begin{equation}
\label{eq:ble112twoSmall}
d_{two} = 0.389 T_c + 0.484 ms
\end{equation}
The mean square error in our measurements for Equation \refeq{eq:ble112twoSmall} is $6.322 \mu s$, the standard deviation $\sigma$ is $2.527 ms$, the maximum deviation that occurred is $5.140 ms$ ($51.4 \%$ of the corresponding measured value).
These values are valid for BLE112-devices in piconets with one slave. For other situations or different hardware, this measurement has to be performed again using the procedure we described. It would be desirable if the manufacturers of BLE stacks would publish data concerning the values of $d_{two}$ and $d_{tw}$ in future, as this information might be important for online power management algorithms on BLE devices.

The connection establishment procedure is always associated with advertising and scanning. Therefore, the charges $Q_{ev,cR,Sl}$ and $Q_{ev,cR,Ma}$ need to be assigned either to the connection establishment model or to the neighbor discovery model.
By definition, we assign it to the neighbor discovery model, accounting for them as the last advertising or scan-event event happening and therefore, set it to zero in the equations described in this section to avoid counting these charges twice.

\subsection{ADV process duration}
\label{sec:advScanningModel}
Advertising and scanning in Bluetooth Low Energy were described in Section \ref{sec:protocol_fundamentals}.
During these phases, the advertiser (A) is not synchronized with the scanner (S) and the reception of a packet sent to the scanner by A is not guaranteed. Instead, the time when a scanner receives an advertising packet can be described as a stochastic process, and it is likely that the advertiser will have to send a packet more than once before it is received by the scanner. This stochastic nature has a strong impact on the energy consumption of the BLE device in this mode, as a packet has to be sent multiple times and the scanner has to scan until a successful reception is achieved.

In many cases, the energy-consumption for device discovery is dominated by the energy spent in advertising or scanning mode without a successful reception. Therefore, the number of advertising events $\overline{N_{a}}$ and scan events $\overline{N_{s}}$ that take place in average before a successful reception occurs have to be calculated. To do this, we derive the average duration of this process $\overline{d_{adv}}$. It is defined as the time from the first advertising packet being sent until there is a successful reception at the scanner and is closely related to the number of events\footnote{We assume that the scanner has already been scanning when the advertiser starts advertising.}.

In this section, we first present a probabilistic model for estimating $\overline{d_{adv}}$. It depends on the protocol parameters $T_a$, $T_s$, $d_s$ and the duration of one advertising packet $d_{advPkg}$, which is calculated in Section \ref{sec:singleEventModel}.
Recall from Section \ref{sec:ble} that the scanner hops between all the three advertising channels and stays on one channel for each scan window $d_s$, whereas the advertiser may freely choose a set of advertising channels to send its packets on.
In this section, we assume that the advertiser sends packets on all three advertising channels. Results for cases when only a subset of channels are used can be calculated in a similar manner.

\subsubsection{Problem description}
Figure \ref{fig:advScanning} shows a typical advertising/scanning situation. A scanner \emph{S} starts listening at its first scan event at time $t=0$. An advertiser \emph{A} begins advertising after a random delay $\phi$ relative to the beginning of the first scan event. It periodically repeats its advertising event with period $T_{a} = T_{a,0} + \rho$, whereby $\rho$ is a random offset between $0 ms$ and $10 ms$. The relevant question that needs to be answered is the following. How much time will pass, until an advertising event ``meets'' a scan event for the first time, or in other words, what is the earliest time when the two events coincide?

A condition for successful reception needs to be identified. An advertising event is received by the scanner, if its starting time $t_a$ is contained within a set of suitable times $t_{i,suc}$. On first glance, $t_{i,suc}$ appears to contain all points in time where the advertising time begins while the node is scanning. Indeed, an advertising event might start even earlier than the subsequent scan event, to be received successfully. In addition, some points in time within a scan event lead to a lost advertising packet. 
The hatched areas in Figure \ref{fig:advScanning} on the left side of the scan events show the possible times $d_{early}$ an advertising event can begin before the scan event starts to be received successfully. The hatched areas at the right sides of the scan events show the times $d_{late}$ that do not lead to a successful reception while a scan event takes place. This is due to each advertising event consisting of three packets on three different channels. $t_{i,suc}$ is defined by the necessary and sufficient condition that the advertising packet on the channel the scanner is scanning on must overlap completely with a scan event. If an advertising packet has the duration $d_{a}$ and changing the channel takes $d_{ch}$ amounts of time, the interval $t_{i,suc}(t_{sE})$ for a given scan event at time $t_{sE}$ can be described as follows.
The set of times $t_{i,suc}(t_{sE})$ for each scan event starts $d_{early}$ time units before the time the scan event begins $t_{sE}$ and ends $d_{late}$ time units before the time the scan event ends $t_{sE} + d_s$. $d_{early}$ and $d_{late}$ depend on the channel of the scan event and are given in Table \ref{tab:tEarlyLate}. This has first been presented in \cite{liu:12_techrep}. As can be seen easily, the effective scan window $d_s' = d_s + d_{early} - d_{late}$ is constant for all channels.

\ifACM
\begin{table}
\tbl{$d_{early}$ and $d_{late}$ and the effective scan window $d_s'$  for the three advertising channels \cite{liu:12_techrep}. $d_{a}$ is the duration of an advertising packet, $d_{ch}$ the duration of hopping to the next channel, as shown in Figure \ref{fig:advScanning}.}{

\begin{tabular}{|c|c|c|c|}
\hline \rule[-2ex]{0pt}{5.5ex} \textbf{Channel} & \textbf{$d_{early}$} & \textbf{$d_{late}$} & \textbf{$d_s'$} \\ 
\hline \rule[-2ex]{0pt}{5.5ex}  Ch 37& 0 &  $d_{a}$ & $d_s - d_{a}$ \\ 
\hline \rule[-2ex]{0pt}{5.5ex}  Ch 38& $d_{a} + d_{ch}$ & $2  d_{a} + d_{ch}$ &  $d_s - d_{a}$ \\ 
\hline \rule[-2ex]{0pt}{5.5ex}  Ch 39& $2 d_{a} + 2 d_{ch}$  &$ 3 d_{a} + 2  d_{ch}$  &  $d_s - d_{a}$ \\ 
\hline 
\end{tabular}}
\label{tab:tEarlyLate}
\end{table}
\else
\begin{table}
\begin{tabular}{|c|c|c|c|}
\hline \rule[-2ex]{0pt}{5.5ex} \textbf{Channel} & \textbf{$d_{early}$} & \textbf{$d_{late}$} & \textbf{$d_s'$} \\ 
\hline \rule[-2ex]{0pt}{5.5ex}  Ch 37& 0 &  $d_{a}$ & $d_s - d_{a}$ \\ 
\hline \rule[-2ex]{0pt}{5.5ex}  Ch 38& $d_{a} + d_{ch}$ & $2  d_{a} + d_{ch}$ &  $d_s - d_{a}$ \\ 
\hline \rule[-2ex]{0pt}{5.5ex}  Ch 39& $2 d_{a} + 2 d_{ch}$  &$ 3 d_{a} + 2  d_{ch}$  &  $d_s - d_{a}$ \\ 
\hline 
\end{tabular}
\label{tab:tEarlyLate}
\caption{$d_{early}$ and $d_{late}$ and the effective scan window $d_s'$ for the three advertising channels \cite{liu:12_techrep}. $d_{a}$ is the duration of an advertising packet, $d_{ch}$ the duration of hopping to the next channel, as shown in Figure \ref{fig:advScanning}.}
\end{table}
\fi
A probabilistic model has to predict the expected number of advertising events before one of them begins within $t_{i,suc}$ and is hence received successfully. The points in time the advertising events start are not independent of each other. Because of this, in general the offset between advertising and scan event can differ for every period of the advertising or scanning process and each period needs to be taken into account. To the best of our knowledge, no closed-form solution exists. We will first give an algorithm that can compute a good approximation for all sets of parameters allowed by the Bluetooth specification. In Section \ref{sec:verification}, we will compare the algorithm's results both with simulation and measurement results.

For the special cases of constant scanning with $d_{s} \approx T_{s}$ and for $T_{a} < d_{s}$, analytical solutions have been described in \cite{liu:12_short,liu:12_long,liu:12_techrep}, which we also describe in this paper for the sake of completeness of description.

\subsubsection{Numeric Approximation}

In general, the problem of calculating the expected advertising latency $d_{adv}$ is a complex stochastic problem. Except for the special cases described above, to the best of our knowledge, no closed-form solution is known. Therefore, we propose an algorithm which estimates the actual results in close proximity, as we will show in Section \ref{sec:verification}.

Algorithm \ref{alg:advScanDur} provides an approximate value of the expected discovery time $\overline{d_{adv}}$. We explain this algorithm below.
Let us start with a given phase offset $\phi$. For this $\phi$, we look at each advertising event $n$ and calculate its probability for a successful reception $p_{hit}(n)$, given that all previous events have not been received at the scanner.
Without the random advertising delay $\rho$, the probability would be estimated to either 0 or 1 - if the advertising begins within $t_{i,suc}$, it is 1, for all other cases, it is 0. With a random offset $\rho$, the time an advertisement event begins, $t_{advEvt}$, can lie within a broad time interval which widens for increasing event numbers $n$. Therefore, $p_{hit}$ depends on $n$. $\rho$ is modeled as a random variable $\rho$, having the distribution
\begin{equation}
\label{eq:uniformDist}
f(\rho) = \begin{cases}
\frac{1}{10 ms} & \textbf{if } 0 \leq \rho \leq 10 ms \\
0				& \textbf{, else}
\end{cases}
\end{equation}
As the Bluetooth specification \cite{bleSpec} only specifies the possible range of values for $\rho$ but not its distribution, this is a further assumption. As $\rho$ is generated by a digital random number generator (RNG) and digital RNGs commonly produce uniformly distributed values, this assumption should hold true for most devices.
The time an advertising event begins $t_{advEvt}$ can be modeled as in Equation \refeq{eq:advEvt}:
\begin{equation}
\label{eq:advEvt}
t_{advEvt}(n,\phi) = \underbrace{\phi + n T_a}_{A} + \underbrace{\sum_{1}^n}_{B} \rho
\end{equation}
For calculating the probability $p_{hit}$ of weather an advertising event is successfully received, the probability density function (PDF) of the start of an advertising event over time $t_{advEvt}(n,\phi)$ is required.
In Term \textit{B} in Equation \refeq{eq:advEvt}, the shape of the distribution depends on n. For n = 1, the distribution is uniform as assumed. As $n$ random offsets $\rho$ occur, the resulting distribution can be described as the sum of $n$ independent and identical random variables $\rho$. In general, the PDF of a sum of n random variables is the convolution product of their PDFs \cite{yates:05}. According to the central limit theorem for large n, the convolution product passes into a Gaussian distribution having the effective mean $\mu' = n \mu$ and the effective standard deviation $\sigma' = n \sigma$ \cite{yates:05}. 
Accordingly, for the distribution $f(B)$ of B, we assume a Gaussian distribution with the mean $\mu_{\rho} = n \cdot 5 ms$ and standard deviation $\sigma_{\rho} = \sqrt{\frac{n}{12}} \cdot 10 ms$ for an arbitrary $n > 2$. For n = 1, the uniform distribution from Equation \refeq{eq:uniformDist} is used. For n = 2, we assume the distribution to be a symmetric triangular distribution with $f(B) = f(\rho) * f(\rho)$ (the exact solution of the convolution product). For the ease of exposition, in this section, a Gaussian distribution is assumed for $n = 1$ and $n = 2$, too. However, the values we present in this paper have been calculated without using this simplification.
To further increase the accuracy of the results, one may pre-calculate the convolution functions for the first $n$ advertising events and use the Gaussian distribution only for large values of n\footnote{Our experiments have shown that using Gaussian PDFs for $n > 2$ provides sufficient accuracy.}.

The Term \textit{A} in Equation \refeq{eq:advEvt} causes a shift in the mean of $f(p_{hit})$. The approximate distribution for the time $t_{advEvt}(n,\phi)$ is:
\begin{equation}
f(t_{advEvt}(n,\phi)) = \Phi(\frac{\phi - n T_a - n \cdot 5 ms}{\sqrt{\frac{n}{12}}\cdot 10 ms})
\end{equation}
$\Phi(t)$ is the cumulative distribution function  of the standard normal distribution.
Therefore, $p_{hit}$ can be calculated as in Line \ref{acl:asd:erf} in Algorithm \ref{alg:advScanDur} by evaluating $f(t_{advEvt}(n,\phi))$ for all scan intervals $k$ the beginning of the advertising event $t_{advEvt}(n)$ might lie within. The first scan event that might receive the advertising event $n$ is given by $k_{min}$ and the last possible one is scan event number $k_{max}$. The indices $k_{min}$ and $k_{max}$ can be calculated with the formulas in Line \ref{acl:asd:kmin} of the algorithm.
With $t_{ai}$ being the ideal point in time an advertising-event starts (i.e. without the random advertising delay $\rho$), $p_{k}$ in Line \ref{acl:asd:erf} can be calculated.
$p_{k}(k,n, t_{ai},d_{early},d_{late}, T_s)$ is the probability for the advertising event $n$ being received successfully by the scanner. It can be calculated as follows.

\begin{align}
p_{k} &= \Phi(\frac{k T_s + d_s - d_{late} - t_{ai} - n \cdot 5 ms}{\sqrt{\frac{n}{12}} \cdot 5 ms}) - \Phi(\frac{k T_s + d_{early} - t_{ai} - n \cdot 5 ms}{\sqrt{\frac{n}{12}} \cdot 5 ms})
\end{align}

Thus, the expected advertising duration for a given offset $\phi$ can be calculated as in Line \ref{acl:asd:texp} of the algorithm.
$p_{cM}(n)$ is the probability that $n$ advertising events do not lead to a successful reception (cumulative miss probability).
In Line \ref{acl:asd:pcM} of Algorithm \ref{alg:advScanDur},  the probability for $n$ cumulative misses is calculated for the current advertising event.
$d_{exp}$ in Line \ref{acl:asd:texp} is a good approximation of the exact\footnote{Exact means the value under all assumptions made so far.} expected advertising duration, which would be:
\begin{equation}
\label{eq:text_exact}
d_{exp} = \int\limits_{k T_s - d_{early}}^{k T_s + d_s - d_{late}}  \Phi(t,t_{ai} + n\frac{10ms}{2},  \sqrt{\frac{n}{12}} 10ms)
\end{equation}
since this equation is computationally complex and the benefit in terms of accuracy is small, the expected value of $\rho$ (i.e., $5 ms$) is used in the algorithm and the error can be neglected.

With increasing values of n, the probability that one of the advertising events considered until so far is received successfully grows. Consequently, $p_{cM}$ shrinks with growing $n$. 
The algorithm finishes, if $(1 - p_{cM})$ is smaller than a lower bound $\epsilon$. The higher $\epsilon$ is, the better the accuracy of the algorithm becomes, but the  corresponding computational complexity increases.

Following the steps described above, the resulting values of $d_{exp}$ must be integrated over all possible values of $\phi$.  We perform a simple numerical integration by evaluating the integral for $\phi \in [0, 3 \cdot T_s]$ in steps of $\Delta$, multiplying the results with $\Delta$ and computing the sum of these values. 
The variable \emph{ch} contains the value of the current channel the scanner receives on and is calculated in Line \ref{acl:asd:determine_channel} of the algorithm.
The function \mbox{\textit{getInterval(ch)}} used in the algorithm looks up $d_{early}$ and $d_{late}$ from Table \ref{tab:tEarlyLate}.
The function \textit{$d_{advEvnt}(ch)$}  calculates the duration of an advertising event which is received successfully by the scanner on the current channel as described in Section \ref{sec:conAdvEvent}. For channel 37 this is $d_{a}$, for channel 38 it is $2 d_{a} + d_{ch}$ and for channel 39 it is $3 d_{a} + 2 d_{ch}$.
 
\algsetup{indent=0.5em}
\begin{algorithm}
\caption{Calculation of advertising duration $d_{adv}$ for case 2b}
\label{alg:advScanDur}
\begin{algorithmic}[1]
\STATE $d_{adv} \gets 0$
\FOR{$\phi = 0$ to $3 T_s$ step $\Delta$}	\label{acl:asd:phi}
 \STATE	$n \gets 0$, $d_{exp} \gets 0 $, $p_{hit} \gets 0$, $p_{cM} \gets 1 $, $ch \gets 37$
  \WHILE{$1 - p_{cM} \le \epsilon$}
   \STATE $t_{ai} \gets \phi + n T_a$,	$k_{min} \gets \lfloor \frac{t_{ai}}{T_s}\rfloor$	\label{acl:asd:kmin}, $k_{max} \gets \lfloor \frac{t_{ai} + n \cdot 5 ms}{T_s}\rfloor$	\label{acl:asd:kmin}
   \STATE $p_{hit} \gets 0$
   \FOR{$k = k_{min}$ to $k_{max}$}								\label{acl:asd:kmax}
    \STATE $ch \gets mod(j,3) + 37$				\label{acl:asd:determine_channel}
     \STATE $(d_{early}, d_{late}) \gets getInterval(ch)$
     \STATE $p_{hit} \gets p_{hit} + p_{k}(k,n, t_{ai},d_{early},d_{late}, T_s)$							\label{acl:asd:erf}
     \STATE $d_{exp} \gets d_{exp}  + p_{k} \mbox{ } p_{cM} \cdot (n \cdot (T_{a} + 5 ms) + d_{advEvnt}(ch))$		\label{acl:asd:texp}
  \ENDFOR
  \STATE $p_{cM} \gets p_{cM}  (1 - p_{hit})$	\label{acl:asd:pcM}, $n \gets n + 1$
\ENDWHILE
\STATE $d_{adv} \gets d_{adv} +  d_{exp}$
\ENDFOR
\STATE $\overline{d_{adv}} \gets \frac{d_{adv}}{\lfloor\frac{3 T_s}{\Delta}\rfloor} $		\label{acl:asd:result}

\end{algorithmic}
\end{algorithm}

In order to bound the computation time, the inner while-loop of Algorithm \ref{alg:advScanDur} should be aborted if $d_{exp}$ exceeds an
upper bound $d_{exp,max}$. In practice, this is not a real limitation because parameterizations leading to advertising durations larger than, for example, $d_{exp,max} = 1000s$ are no reasonable choices for real world applications.
The long latencies resulting from such parametrizations cannot be accepted and are not pareto-optimal.
The algorithm has two parameters $\epsilon $ and $\Delta$ that influence the quality of the result. The smaller $\Delta$ is and the closer $\epsilon$ gets to 1, the more accurate the result becomes.
Compared to discrete event simulations, this algorithm has a reduced complexity when choosing these values appropriately, as not all scan events need to be examined, and as different values for $\rho$  are accounted for by the Gaussian distribution which is used by the algorithm.

We assume $d_{a}$ to be equal to $446 \mu s$, which is the time spent for sending a 37-byte advertising packet plus the interframe space, neglecting the small additional time for idle-listening. This value can also be used for the last advertising packet that receives a response and hence lasts slightly longer than the packets before, as the error introduced by this is low. $d_{ch}$ is assumed to be $d_{IFS} = 150 \mu s$.
\begin{figure}[htb]
\centering
\includegraphics[width = 13cm]{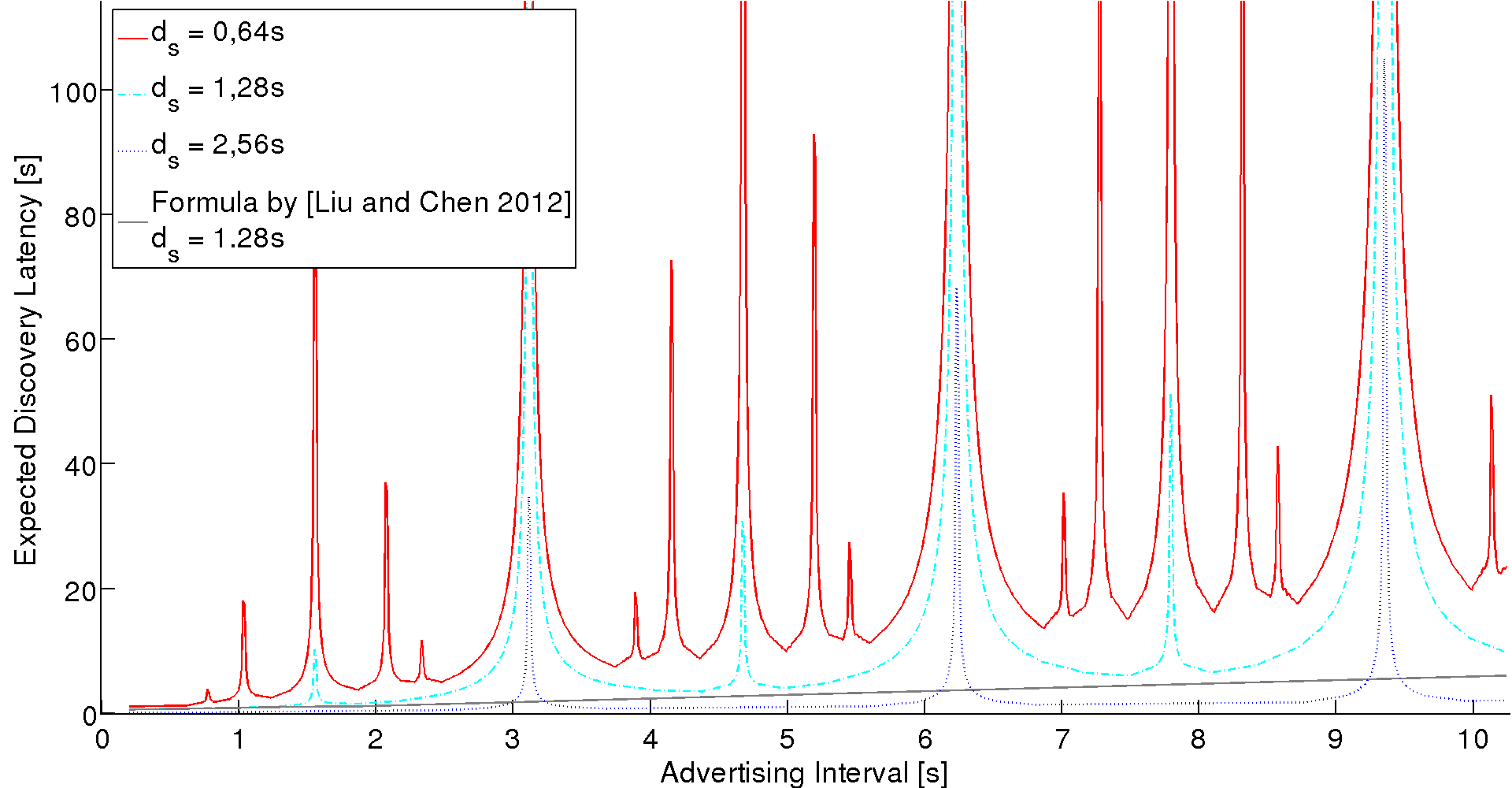}
\caption{Results of the algorithm for $T_s = 3.12,\mbox{ } \epsilon = 0.9999,\mbox{ }\Delta = 93.6 ms,\mbox{ }d_a = 446 \mu s,\mbox{ }d_{ch} = 150 \mu s $. The black, solid line shows the results of the analytical model presented in \cite{liu:12_long} for $T_a < d_{s}'$ with $T_s = 2.56 s$ and $d_{s} = 1.28 s$.}
\label{fig:modResultVarIa} 
\end{figure}

Figure \ref{fig:modResultVarIa} shows the result for the expected advertising duration over $T_a$ for different scan windows $d_s$.
It leads to some interesting observations:
\begin{itemize}
\item Smaller values of $d_s$ lead to higher advertising latencies $\overline{d_{adv}}$ for all values of $T_{a}$. 
\item For $T_a < d_s'$, the function rises only slowly and there are no peaks.
\item For some parameter valuations (for example, for $T_a = n\mbox{ } T_s$) the expected advertising latency $d_{adv}$ becomes very large. These are coupling phenomena. If the first advertising event does not hit a scan event and its advertising interval $T_a$ is for example close to the multiple of the scan interval $T_s$, the next advertising events have a high probability of missing the scanner too.
This coupling phenomenon has also been detected by simulations in \cite{liu:12_long} for $T_s \approx T_a$. Obviously, the 10 ms of random advertising delay $\rho$ are not enough to overcome this if the gap between the scan events $T_s - d_s'$ is large enough. For small scan windows $d_s$, this phenomenon intensifies and the number of coupling-peaks increases.
\item Some coupling-phenomena have peaks with smaller magnitudes, for example for $T_a = \frac{3}{2} T_s$ in the curve for $d_s = 1.28s$.
\item In contrast, local minima exist that lead to short advertising latencies $\overline{d_{adv}}$.
\end{itemize}

We now discuss some special cases for notable parameter valuations.

\subsection{Special cases}
Recently, closed-form models for some of the special cases have been published. For completing our descriptions, we present these models below using the same symbols and notations as presented in our paper.
Whereas the estimations of Algorithm \ref{alg:advScanDur} on the discovery-latency involve some computational complexity, these models gives precise values with no noteworthy computational complexity. 
However, these models are limited to the special cases $T_s \approx d_s'$ and $T_{a} <= d_s'$.\\
\textbf{Continuous scanning: $d_s \approx T_{s}$}\\
Parametrizations with $d_s \approx T_{s}$ are known as \emph{continuous scanning} \cite{liu:12_long} as the scanner continuously scans without sleeping between the scan events. This situation leads to very low discovery latencies but the scanner consumes more energy when it has to wait for the first advertising packet being sent for a long time (idle-scanning). For this situation, an analytical model that has been presented in \cite{liu:12_long}, which we describe below. 

\begin{figure}[b]
\centering
\includegraphics[width = 8cm]{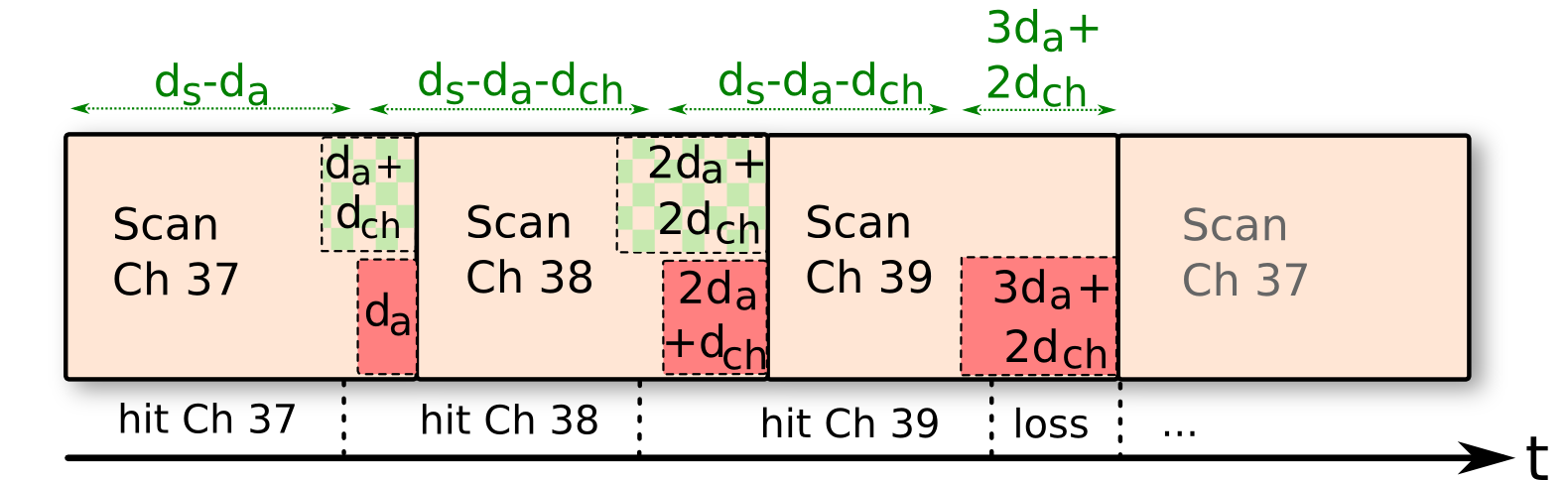}
\caption{Scan events with continuous scanning}
\label{fig:advScanningCont} 
\end{figure}

In continuous scanning, the  scanner scans continuously on all three channels, hopping to the next channel after the time $d_s \approx T_s$ has passed. Thereby, we assume that there is no time needed for changing the channel\footnote{This assumption can be made as the time for channel-changing is small. For a BLE112, we measured an average channel changing duration $d_{chCh} = 1,325ms$.}.
Figure \ref{fig:advScanningCont} depicts the different time intervals an advertising event can begin within. The hatched boxes depict $d_{early}$ and the solid boxes $d_{late}$ for each channel. Except for channel 39, $d_{late}$ overlaps with $d_{early}$ of the adjacent channel. An advertising packet is received successfully if it begins at a point in time the scanner is scanning, except its beginning is up to $d_{late}$ time units before the scanner performs a channel change. In addition, it is also received successfully $d_{early}$ time units before the scanner starts scanning on a channel. As a consequence, the only region that can lead to a loss of the advertising packet is $d_{late}$ before a channel change from channel 39 to 37. In this case, the packet has to be repeated. Other regions always lead to a successful reception on the channel as written at the bottom of Figure \ref{fig:advScanningCont}.

As can be seen in Figure \ref{fig:advScanning}, the advertiser sends its first advertising packet with a random offset $\phi$ after the beginning of the first scan event. The offset $\phi$ is a uniformly distributed 
random variable and hence each offset occurs with the same probability. If $\phi$ is larger than $3 T_s$, the situation in Figure \ref{fig:advScanningCont} repeats periodically, therefore it is sufficient to describe the problem for  $\phi \in [0..3 T_s]$.

If $T_a \le 3 T_s - 3 d_a - 2 d_{ch}$, the duration until reception is bounded. The maximum duration $d_{adv,max}$ is:
\begin{equation}
d_{adv,max} = T_{a} + 3 d_{a} + 2 d_{ch}
\end{equation}

\begin{figure*}[t]
\centering
\includegraphics[width = 13cm]{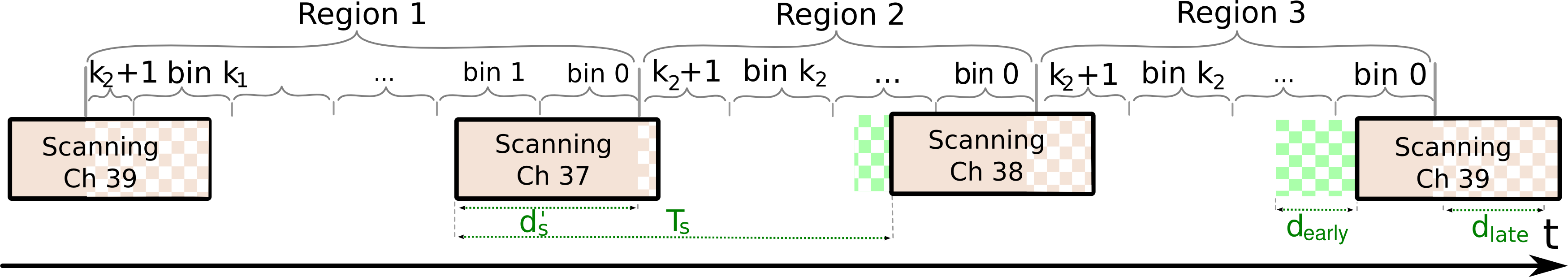}
\caption{Model for the bounded latency case (graphic based on \cite{liu:12_short})}
\label{fig:nokiamodel} 
\end{figure*}

A solution for the average duration $\overline{d_{adv}}$ has been described in \cite{liu:12_long}. In this paper, we use slightly different interval boundaries compared to \cite{liu:12_long}. Whereas the latter assumes the same hit probabilities for all three advertising channels, we assign a slightly higher hit probability to channel 37 due to its wider success-interval $t_{i,suc}$, as can be seen in Figure \ref{fig:nokiamodel}. We believe that the ones according to those in Figure \ref{fig:advScanningCont} are more precise. In addition we account for the random delay $\rho$ by using an effective advertising interval $T_a' = T_a + 5 ms$. However, the resulting values for our modifications and the original model in \cite{liu:12_long} lie in close proximity. The hit-probabilities on a particular channel $P_{chan}$ and the loss-probability $P_{loss}$ are:
\begin{align}
P_{37} = \frac{d_s-d_a}{3 d_s} & & P_{38} = P_{39} = \frac{d_s-d_a - d_{ch}}{3 d_s} & &P_{loss} = \frac{3 d_a + 2 d_{ch}}{3 d_s} 
\end{align}

The average advertising duration $\overline{d_{adv}}$ is:
\begin{align}
\overline{d_{adv}} & = d_{a} P_{37} + (2 d_{a} + d_{ch}) P_{38} + (3 d_{a} + 2 d_{ch}) P_{39} + (T_a' + d_{a}) P_{loss}  = \nonumber \\
				& =  \frac{2 d_{ch} T_{a}'}{3 d_s} + \frac{d_a T_{a}'}{d_s}  - \frac{d_{ch}^2}{d_s} - \frac{2 d_a d_{ch}}{d_s} - \frac{d_a^2}{d_s} + d_{ch} + 2 d_a
\label{eqn:d_adv_contscan_1}
\end{align}	

Even though a coupling phenomenon could occur for $T_a = T_s$, we could never observe any in our simulations. Therefore, this formula gives a good approximation also for the case $T_a > d_s'$.
\\
\textbf{Bounded latency:} $d_s < T_{s}$ and $T_{a} <= d_s'$\\
In this case, the latency is bounded because an advertising event is guaranteed to be received within 1 advertising interval as the scanner is constantly scanning for the time in between two advertising events and no coupling-phenomenon can occur. The maximum discovery latency is
\begin{equation}
d_{adv,max} = \lceil\frac{T_s - d_s}{T_a}\rceil T_a + 3 d_a + 2 d_{ch}
\end{equation}

For a similar problem in the context of the protocol STEM-B, a deviation of the average discovery latency has first been given in \cite{schurgers:02} and has been applied to Bluetooth Low Energy in \cite{liu:12_short,liu:12_long,liu:12_techrep}. This solution is depicted in Figure \ref{fig:nokiamodel} and described below. It is only valid for $T_a < d_s'$.
Similar to the solution described for Algorithm \ref{alg:advScanDur}, the first advertising event happens at a random time with a maximum phase offset $\phi = 3 T_s$ from the beginning of a scan event on channel 37. The value of $\phi$ determines the advertising latency $d_{adv}$ together with the random \mbox{offset $d_{adv}$} which only has a small impact on the latency. If an advertising event starts within one of the three regions in Figure \ref{fig:nokiamodel}, the channel an advertising packet is received on is determined and thus, the duration of the last advertising event is determined too. In addition, there are k + 2 sections in each region. The section decides, how many advertising intervals take place before a successful reception occurs.
Using the ideas described above, a formula can be derived.

The result for the discovery latency $\overline{d_{adv}}$ presented in \cite{liu:12_long} and evaluated for $N = 3$ advertising channels is:
\begin{align}
\label{eq:nokiamodel}
\overline{d_{adv}} &= (d_a + \frac{T_s-T_a}{2}) - \frac{(d_a + d_{ch}) (d_a - d_s)}{T_s} - \frac{1}{3} h_2(1) \cdot (1-\frac{T_a + h_2(1)}{2T_s})\nonumber\\
 & + \frac{1}{3} h_2(2) \cdot (1-\frac{T_a + h_2(2)}{2T_s}) + \frac{1}{3} h_2(3)\cdot(1-\frac{T_a + h_2(3)}{2T_s}) +  \frac{T_a^2}{2 T_s} ( h_1(1)\cdot(1-h_1(1))\nonumber\\
 & + h_1(2) \cdot (1-h_1(2)) + h1(3) \cdot (1-h3(3)))  + h_3 
\end{align}

with:
\begin{align}
\begin{array}{lcl}
h1(n) = \frac{(A(n) + T_s) - (B(n) + d_s)}{T_a}  - C(n) & & h2(n) = (d_s - T_a) + (B(n) - A(n))  \nonumber\\
h3 = \frac{T_a  (d_a + d_{ch}) (h_1(2) + 2 h_1(3)))}{3 T_s} & & A(n) = n d_a  + (n-1) d_{ch} \nonumber\\
B(n) = A(n)+ d_{ch} & & C(n) = \lfloor\frac{(A(n) + T_s) - (B(n) + d_s))}{T_a} \rfloor
\end{array}
\end{align}
\begin{figure}[b]
\centering
\includegraphics[width = 13cm]{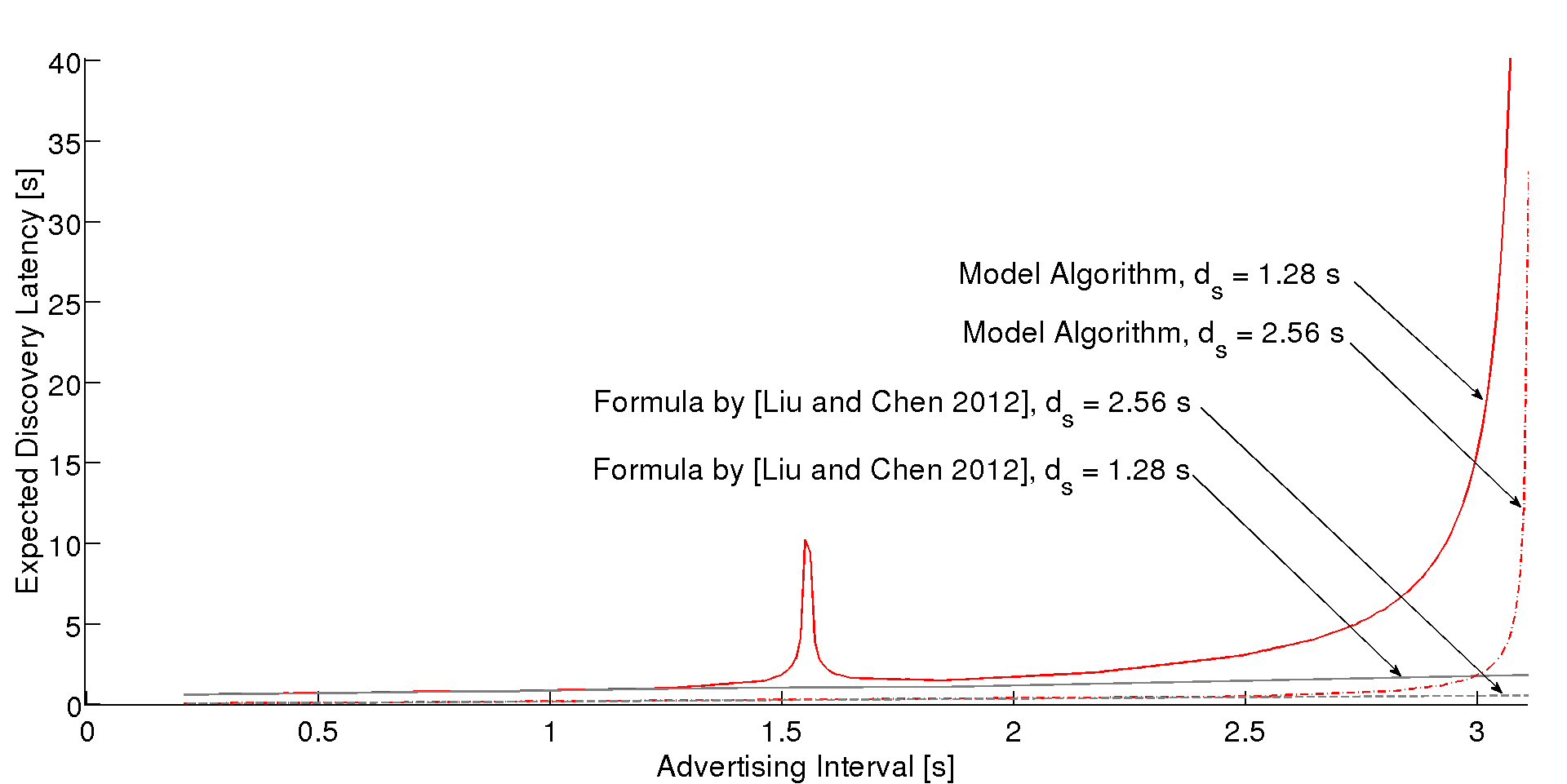}
\ifACM
\caption{Model from Equation \refeq{eq:nokiamodel} compared to result from Algorithm \ref{alg:advScanDur} ($T_s = 3.12 s,\mbox{ }\\
\Delta = 93.6 ms,\mbox{ }d_a = 446 \mu s,\mbox{ }d_{ch} = 150 \mu s$)}
\else
\caption{Model from Equation \ref{eq:nokiamodel} compared to result from Algorithm \ref{alg:advScanDur} ($T_s = 3.12 s,\mbox{ } \Delta = 93.6 ms,\mbox{ }d_a = 446 \mu s,\mbox{ }d_{ch} = 150 \mu s$)}
\fi
\label{fig:nokiamodelplot} 
\end{figure}
Figure \ref{fig:nokiamodelplot} shows that Equation \refeq{eq:nokiamodel} produces equal results as Algorithm \ref{alg:advScanDur} as long as $T_a < d_s'$, but cannot predict the discovery latency for $T_a > d_s'$.
\\
\textbf{Unbounded latency: $d_s < T_{s}$ and $T_{a} > d_s'$}\\
In this region, coupling peaks occur as already described. To the best of our knowledge, our model is the first one that can give estimations for this case.
In Figure \ref{fig:modResultVarIa}, it can easily be seen that choosing $T_a$ appropriately for given values of $T_s$ and $d_s$ is crucial for short latencies $d_{adv}$ and for a low energy consumption during device discovery.
As there is no closed-form formula, a good suggestion on how to choose the right parameter values is making use of Algorithm \ref{alg:advScanDur} to generate a plot of possible values
and making an appropriate choice based on the outcome of the algorithm. Embedded power managers can optimize the protocol parameters either by implementing the model or by using a previously computed table of model values for the parameters relevant for the application.

\newpage
\subsection {Computing the number of events}
After having presented a model for each type of event that can occur and a model for the expected advertising latency $\overline{d_{adv}}$, 
we combine both to obtain the energy consumption for device discovery. Therefore, we calculate the number of events that take place within $d_{adv}$ to obtain the energy consumption of the advertiser and the scanner. For the connected mode, this problem is much simpler as the repetition period of the connection events is known and the number of events that occur within a given amount of time can be easily calculated.
\label{sec:noEvnentsCalc}

\subsubsection{Connected Mode}
\label{sec:3rdCon}
For a master, the number of connection events $N_{c,ma}$ in a given amount of time $T_{g}$ is
\begin{equation}
N_{c,ma} = \lfloor \frac{T_{g}}{T_c} \rfloor
\end{equation}
For the slave, the number of connection events $N_{c,sl}$ is different as it may skip some events due to the slave latency parameter $N_{sl}$. With an average number of $\overline{N_{sl}}$ skipped events, the energy-consumption for the slave is
\begin{equation}
N_{c,sl} = \lfloor \frac{T_{g}}{\overline{N_{sl}} T_c} \rfloor
\end{equation}

The overall charge consumption of the master or the slave is the sum of the event energies and the sleeping energy:
\begin{align}
\label{eq:pwrConsumptionMaster}
\overline{Q_{con,Ma/Sl}} &= \sum\limits_{n=1}^{N_{c}}Q_{event}(n)  + (T_{g} - \sum\limits_{n=1}^{N_{c}}d_{event}(n)) I_{sl}
\end{align}

In Equation \refeq{eq:pwrConsumptionMaster}, $N_c$ can be set to the number of connection events for the master ($N_c = N_{c,ma}$) or for the slave ($N_c = N_{c,sl}$). $Q_{event}$  can be calculated according to \mbox{Section \ref{sec:singleEventModel}}. $I_{sl}$ is the sleep current of the device and $d_{event}(n)$ the duration of event $n$.

\subsubsection{Advertiser}
\label{sec:3rdAdv}
The energy consumption of one advertising event $Q_{advEvent}$ can be modeled as described in Section \ref{sec:singleEventModel}.
For calculating the expected energy consumption  $\overline{Q_{adv}}$ for device discovery of the advertiser, it must be taken into account that not all advertising events are identical. Events that are successfully received might be shorter than the other events as packets on different advertising channels are skipped after a successful reception, and the \emph{rx}-phase takes longer due to receiving the response.

The exact calculation can only be done within Algorithm \ref{alg:advScanDur} or, in cases there is an analytical formula available, by modifying this formula. In the case of non-continuous scanning with $T_a < w_s'$, this has been done in \cite{liu:12_techrep}. In general, Algorithm \ref{alg:advScanDur} can easily be extended for an estimation of the advertiser's energy. With $\overline{Q_{adv}}$ being the expected energy of the advertiser, one could add the following assignment after \mbox{Line \ref{acl:asd:texp}} of \mbox{Algorithm \ref{alg:advScanDur}}:
\begin{algorithmic}
\STATE $\overline{Q_{adv}} \gets \overline{Q_{adv}} + p_{k}\mbox{ } p_{cM} ((n- 1) (Q_{full} + (T_a - d_{full}) I_{sl})  + Q_{last}(ch)) $
\end{algorithmic}
$Q_{full}$ is the charge consumed by full advertising packet that is sent on all three channels without receiving a response and $d_{full}$ is the corresponding duration. $Q_{last}(ch)$ is the charge of a packet that is successfully received on channel \textit{ch}. It can be calculated according to Section \ref{sec:singleEventModel}. 

To reduce complexity, we suggest an approximate solution by estimating the energy consumed solely from the expected advertising/scanning duration $\overline{d_{avg}}$. With this approximation, the result for $\overline{d_{adv}}$ computed by the algorithm or any analytic result from one of the formulas presented can be used to calculate the energy consumption.

In the formulas below, $Q_{39}$ is the charge consumed by sending a full advertising event on all three channels and $d_{39}$ is its duration including pre- and postprocessing; $d_{37}$/$Q_{37}$ and $d_{38}$/$Q_{38}$ is the duration/energy of the last advertising event being received successfully on channel 37 or 38, respectively. $I_{sl}$ is the sleep current of the device. We assume that $Q_{39}$ and $d_{39}$ are constant no matter weather a response is received or not, neglecting the small error introduced by this.

We define the expected number of advertising events that have occurred $\overline{N_a}$, as:

\begin{equation}
\overline{N_a} = \overline{d_{adv}}/(T_a + 5ms) - 1
\end{equation}
and the average energy and duration of the last packet
\begin{align}
\overline{Q_{last}} = \frac{Q_{37} + Q_{38} +  Q_{39}}{3} &, & \overline{d_{last}} = \frac{d_{37} + d_{38} + d_{39}}{3} \nonumber
\end{align}

For the expected energy, there are three cases.
If $\overline{d_{adv}} \le \overline{d_{last}}$, then one has:
\begin{equation}
\label{eq:expectedAdvEnergy1}
\overline{Q_{adv,1}} = \frac{\overline{d_{adv}}}{\overline{d_{last}}}\overline{Q_{last}} 
\end{equation}
If  $\overline{d_{last}} \leq \overline{d_{adv}} \le T_a$, then the expected energy is:
\begin{equation}
\label{eq:expectedAdvEnergy2}
\overline{Q_{adv,2}} = \overline{Q_{last}} + (\overline{d_{adv}} - \overline{d_{last}}) I_{sl} 
\end{equation}
and if $\overline{d_{adv}} > T_a$, one has:
\begin{equation}
\label{eq:expectedAdvEnergy3}
\overline{Q_{adv,3}} = N_{a} Q_{39} + \overline{Q_{last}} + (\overline{d_{adv}} - N_a T_a - \overline{d_{last}}) I_{sl}
\end{equation}
If $N_{a} \ge 1$, then an additional term must be added to account for the sleeping duration of unsuccessful advertisement events:
\begin{equation}
\overline{Q_{adv,3}} += (N_a - 1)(T_a - d_{39})I_{sl}
\end{equation}

\subsubsection{Scanner}
\label{sec:3rdScan}
Given $d_{adv}$ from Section \ref{sec:advScanningModel}, the energy consumption of the scanner is easy to calculate.
However, it depends on its definition. We assume that the advertiser starts advertising  with a random offset $\phi$  with a maximum value of $3 T_s$ from the beginning of the fist scan event. The power consumption of idle scanning before the advertiser starts advertising is not accounted for in the equations below.

The expected energy consumption of the scanner $\overline{Q_{s}}$ is:
\begin{equation}
\label{eq:energyScanner}
\overline{Q_{s}} = \overline{Q_{active}} + \overline{Q_{sleep}}\
\end{equation}
with
\begin{equation}
\label{eq:scanningTime}
\overline{Q_{active}} = \overline{N_s} Q_{sEv,idle}(d_s)
\end{equation}

\begin{equation}
\overline{Q_{sleep}} = \overline{N_s} (T_s - d_s) I_{sl}
\end{equation}
and
\begin{equation}
\overline{N_s} = \frac{\overline{d_{adv}}}{T_s}
\end{equation}

$\overline{N_s}$ in Equation \refeq{eq:scanningTime} is the expected number of scan events that occur within $\overline{d_{adv}}$. $\overline{Q_{active}}$ accounts for the time the scanner is actively scanning and $\overline{Q_{sleep}}$ accounts for the sleep current $I_{sl}$.

$Q_{sEv,idle}$ is the energy consumed by a scan event without receiving an advertising packet as described in Section \ref{sec:scanEvent}, neglecting the energy consumed by sending the response.

\subsection{Parameter extraction}
\label{sec:matlab_script}
\begin{figure}
\centering
\includegraphics[width = 9.0cm]{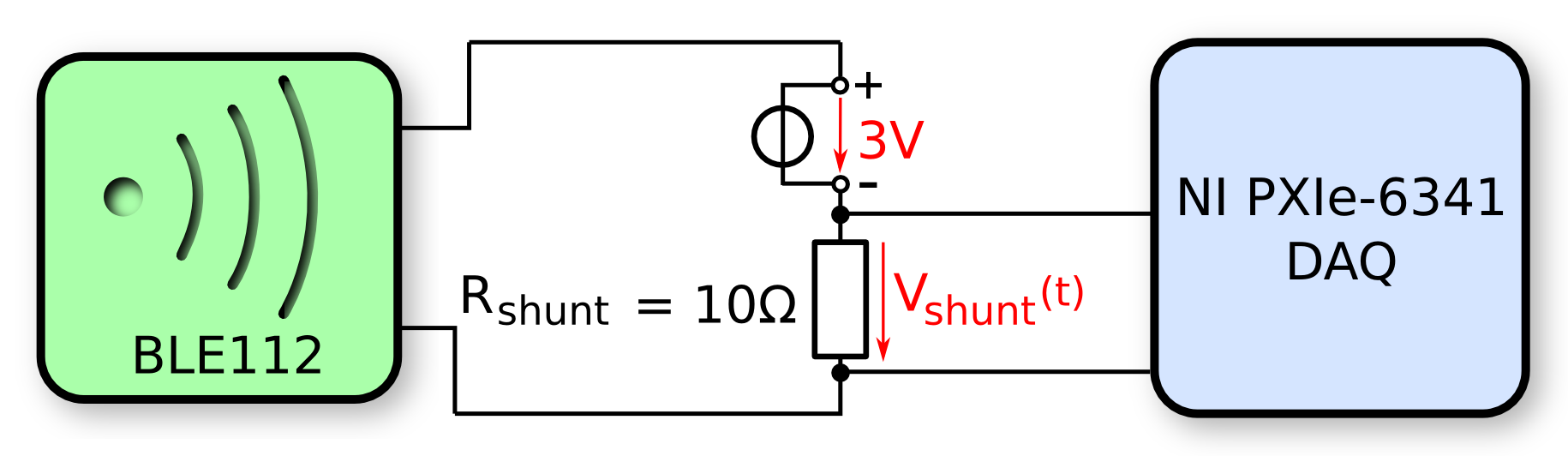}
\caption{Measurement Setup}
\label{fig:measurement_setup}
\end{figure}

To obtain the parameter values needed for the model equations described, a semi-automatic procedure has been used to generate current waveforms and extract the values needed.
Using a Bluegiga firmware created with a custom BGScript \cite{ble112}, we altered the values of important protocol parameters such as the number of bytes sent in a packet and measured the waveforms of more than 5000 connection events.
The measurements were taken according to \cite{tian092:12} using a National Instruments NI PXIe-6124 data acquisition board (DAQ) sampling with a rate of $100 kHz$. We calculated the current consumption $I(t)$ of the chip by analyzing the voltage $V_{shunt}(t)$ measured by the DAQ across a $10\Omega\pm 1\%  $ shunt resistor, as shown in Figure \ref{fig:measurement_setup}. 

The measured waveforms were processed and analyzed using Matlab scripts we developed. The model parameter values we obtained using these scripts are presented in Appendix \ref{sec:appendixParamsConnected} and \ref{table:dataForScanning}.

\section{Verification of the model}
\label{sec:verification}
In this section, we discuss the validity of our proposed model. Towards this, we conducted measurements both in the non-connected and in the connected mode and compare the results with the modeled data.
The only previous comparative measurements of such a model with real-world measurements or simulations have been presented in \cite{liu:12_techrep},\cite{liu:12_short}, \cite{liu:12_long} for device-discovery with $T_a < d_s$.  
In this paper, the discovery latency in the advertising-/scanning mode has been modeled and compared with measured data also for $T_a > d_s$. For the connected mode, we measured the current consumption of a BLE112 module for a given scenario and compared it to the modeled values. The results show that modeled and measured data lie in close proximity.
\subsection{Connected mode}
\begin{figure}[t]
\centering
\includegraphics[width = 13cm]{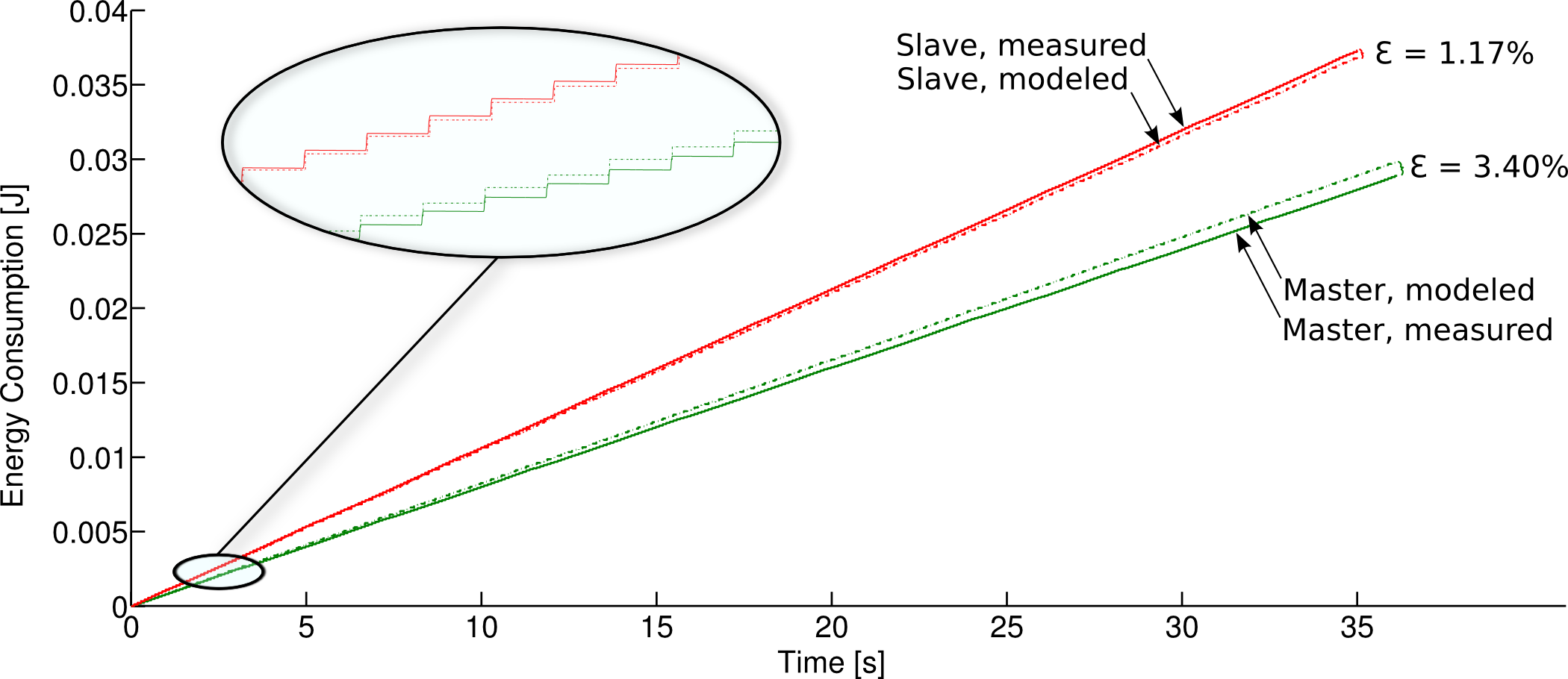}
\caption{Energy consumption for sending a test sequence both modeled and measured. The test sequence consists of packets with varying payload between 0 and 20 bytes. $T_c = 100ms$.}
\label{fig:compSim_connected} 
\end{figure}

To compare the modeled and the measured energy consumption in the connected mode, we set up a test scenario. Two devices exchange data using a connection interval $T_c$ of $100ms$. One device sends data with an attribute-write request, the other node acknowledges the data received. We did this comparison for both, a master and a slave sending the payload. The payload sent varies from 1 to 20 bytes, the number of packets sent per event is one per direction. The number of total bytes received in a response is either 10 or 15, depending on the event. The sequence of connection events in our scenario is as follows: 
\begin{itemize}
\item Send a byte with a payload sweeping between 0 and 20 bytes with 17 bytes overhead\footnote{Sending a zero-byte-packet generates an empty polling packet having a total length of 10 bytes.} and receive an empty polling packet with 10 bytes total length.
\item Send an empty polling packet with 10 bytes length and wait for the acknowledgement packet with 15 bytes length.
 \end{itemize}
 The model uses parameter values according to the tables in Appendix \ref{sec:appendixParamsConnected}. As the device showed an unexpectedly long duration of the first \emph{rx}-phase of each connection event when acting as a slave, we did not use the value of $d_{prerx}$ from the table, but used the effective duration of $d_{prerx} = 388 us$\footnote{This value has been calculated based on our experiments for the values presented in Appendix \ref{sec:appendixParamsConnected} and is valid for a BLE112 device for the first \emph{rx}-phase of a connection event of a slave. We did not find any dependence of this value on the connection interval. A measurement with a CC2540-device with TI's BLE stack did not show such a significant abnormality.} for the slave.

To achieve a hight accuracy, we scheduled the preparation of the attribute-write requests in a task that is run periodically on the BLE module by a separate timer. Consequently, it does not run within the pre- and postprocessing phase of an event. The current waveforms generated by this task were filtered out by a matlab script. In addition, we made sure the BLE device did not send anything other than the packets we took into account in our model, such as a connection parameter update, during the time we were measuring.
In our measurements, even though we used high-quality measurement equipment, the bias current of the measurement device was larger than the sleep current of the BLE device. Therefore, in average, its measured value was slightly negative. 
A Matlab-script detected the time when each connection event occurred in the measured current waveform and triggered the model to compute the charge consumed for the same event at the same point in the simulated time.

Figure \ref{fig:compSim_connected} shows the result of the experiments described. The curves depict the energy-consumption over time for a master and a slave, both computed with our proposed model and obtained by measurements.
As it can be seen, for the slave, the total error of the mean current predicted by our model $\varepsilon_{m} = \frac{\lvert I_{measured} - I_{model} \rvert}{I_{measured}}$ is $1.17 \%$. When only considering the connection-events without the sleep periods in between the events,  the relative error $\varepsilon_{m}$ of the current is $3.5 \%$. If only the parts of an event beginning with the communication preamble and ending with the \emph{tx}-transient phase are considered, $\varepsilon_{m}$ is $ 5.1 \%$. This error is caused mainly by a missprediction of the rx-phase duration. As the error for the whole event is less, some of the error is compensated by the pre- and postprocessing phase. In our test-scenario, we sent the payload and waited for a confirmation from the remote node, whereas we obtained the model-values in Appendix \ref{sec:appendixParamsConnected} for communication without confirmation. The confirmation causes a rise of the average pre- and postprocessing duration. By adjusting these values for sending with confirmation ($d_{pre} = 413 \mu s$, $d_{post} = 1077 \mu s$), the error of the whole event including the sleep current increases to $6,0 \%$, as the error-compensation described does not occur anymore. 

For the master, the overall error of the mean current in the test scenario $\varepsilon_{m}$ is $3.4 \%$. Considering the connection events only, $\varepsilon_{m}$ is $0.6\%$ . Therefore, the error for the  events only is lower than for the slave. The parameter values in the tables of Appendix  \ref{sec:appendixParamsConnected} are only precise for a BLE slave and differ slightly for a master. In particular, $d_{Cpre}$ is longer for the master, but this error is compensated by the pre- and postprocessing phases. Hence, the error is low. As a consequence, the higher overall error must be caused by the sleep current that we could not measure precisely, as described above. For the slave, this error further reduces the overall error by compensating overestimations of the events by the model.

In conclusion, the accuracy reached is sufficient for most applications of the model. Further accuracy could be gained if a separate set of values for BLE masters and different pre- and postprocessing durations for different functions of the BLE device and for different rules (master, slave) would be used.

\ifACM
\else
\FloatBarrier
\fi
\subsection{Advertising/Scanning}
\begin{figure}[t]
\centering
\includegraphics[width = 13cm]{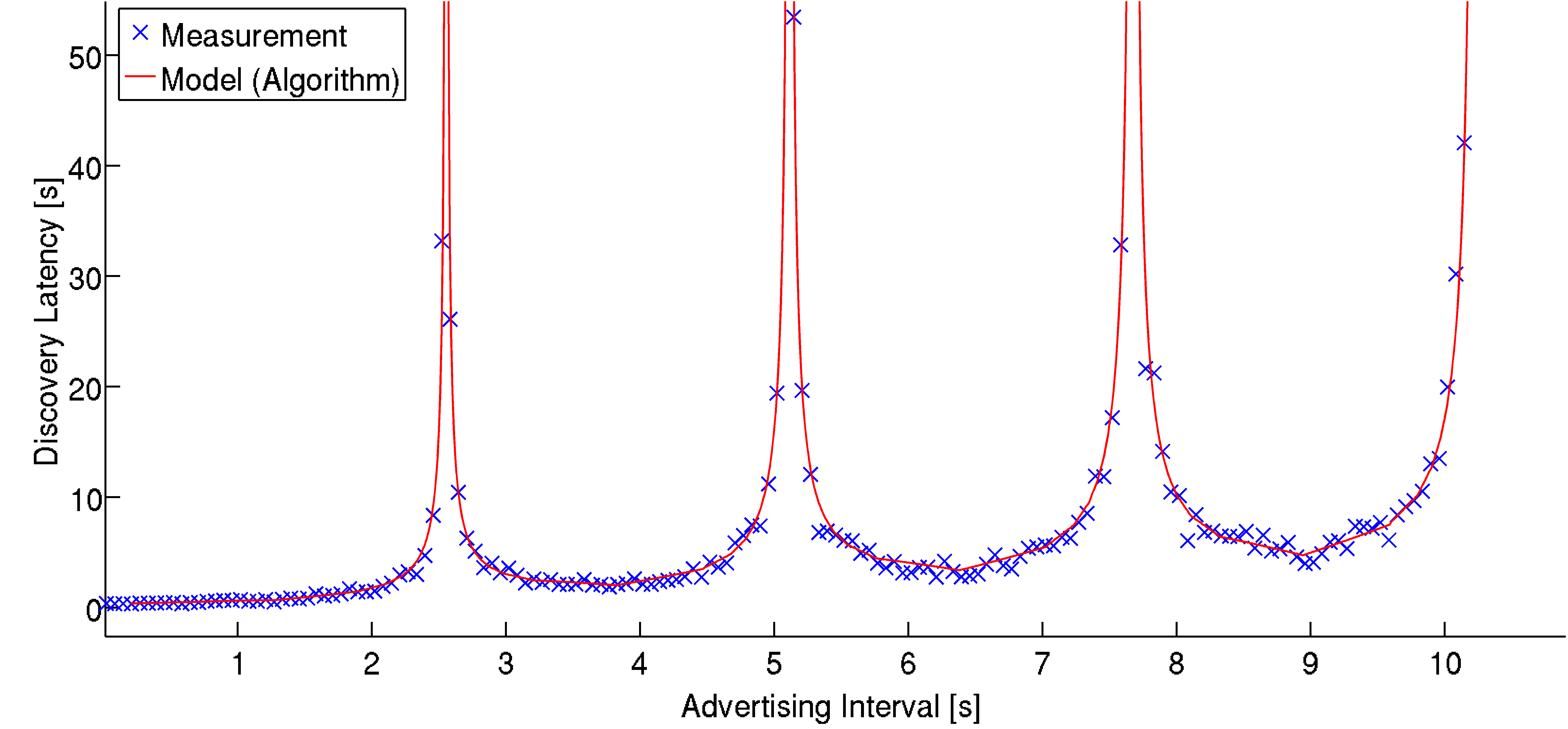}
\ifACM
\caption{Measured discovery latency compared to modeled values. Model parameters: $T_s = 2.56 s,\\d_s = 1.28 s,\epsilon_{m} = 0.9999,\mbox{ }\Delta = 76.8 ms,\mbox{ }, t_{max}=1000s$ (model aborted for times above), $d_{a} = 446 \mu s$, $d_{ch} = 150 \mu s$.}
\else
\caption{Measured discovery latency compared to modeled values. Model parameters: $T_s = 2.56 s, d_s = 1.28 s,\epsilon_{m} = 0.9999,\mbox{ }\Delta = 76.8 ms,\mbox{ }, t_{max}=1000s$ (model aborted for times above), $d_{a} = 446 \mu s$, $d_{ch} = 150 \mu s$.}
\fi
\label{fig:compSim_testsweep_100} 
\end{figure}

For verifying our algorithm, we measured the mean discovery latency $\overline{d_{adv}}$ and compared these results with results from our model as shown in Figure \ref{fig:compSim_testsweep_100}. 

The dashed curve shows the computed advertising latency $d_{adv}$ for different advertising intervals $T_a$. In addition to this curve which has been obtained by using Algorithm \ref{alg:advScanDur},  each cross in the figure shows the mean value of 100 real-world measurements that have been repeated with the same parameter values. These measurements have been performed as follows. Two BLED112-modules have been connected to a PC with the following configuration. One module is scanning continuously with fixed parameters. As soon as a scan response is received, the process controlling this module sends the time of reception to another process controlling the second module. Inter-process communication is handled by Unix-Domain-Sockets. The process controlling the second module starts the advertising procedure at a random point in time between $0s$ and $3 T_s$ and calculates the duration between the start of the advertising and the first reception at the scanner. After having repeated this 100 times for one advertising interval, the average duration is computed and $T_a$ is increased by $63.5 ms$. This sequence is repeated until the maximum advertising interval is reached. Advertising intervals that lie within a coupling peak according to the model have been skipped as the measurement would take a very high amount of time. 
As it can be seen in Figure \ref{fig:compSim_testsweep_100}, results from the measurements and results obtained from the model lie in close proximity. In addition to real-world measurements, we compared results from the model to results from discrete event simulations which also lay in close proximity. As the issue is similar to connected mode and comparative measurements for advertising (for $T_a < d_s'$) have already been done in \cite{liu:12_techrep}, we did not measure the device discovery energy.
\ifACM
\else
\FloatBarrier
\fi
\section{Results and Conclusion}
\label{sec:results_and_conclusion}
\subsection{Sensitivity Analysis}
Some of the durations and current magnitudes of the phases in a connection event vary randomly. For example, the duration of the post-processing phase $d_{post}$ is subject to strong random variations. Another varying value is the current magnitude in the transmission phase, $I_{tx}$.
If the standard deviations and the minimal/maximal values for the varying parameters are known, the impact of these variations on the overall current consumption within one connection interval can be determined. This is important for getting an estimation for the maximum short-time current consumption.

In this section, we analyze the impact of variations of $d_{post}$ and $I_{tx}$ on the energy consumption in one connection interval. Thereby, we assume that the device is sleeping whenever there is no connection event taking place. For other parameters, this sensitivity analysis can be done similarly.
\newline Given a phase $ph$, $d_{min}\le d_{ph}\le d_{max}$, $I_{min}\le I_{ph}\le I_{max}$, the sensitivity $S$ on the duration $d_{ph}$ or current $I_{ph}$ of a phase is:
\begin{align}
S(d_{ph}) &= \frac{\Delta Q_{total}}{\Delta d_{ph}} & S(I_{ph}) &= \frac{\Delta Q_{total}}{\Delta I_{ph}}
\end{align}
$Q_{total}$ is the total charge consumed within one connection interval, as calculated in Equation \refeq{eq:pwrConsumptionMaster} with $N_{c} = 1$.
All parameter values can be assumed to be independent from each other. Therefore, the sensitivity analysis can be done by accounting for the variations of each parameter individually, assuming that all other parameters have fixed values.
With $Q_{sl}$ ($I_{sl}$) being the charge consumed (current drawn by sleeping), the sensitivity $S_{d_{ph}}$ can be written as:
\begin{align}
S(d_{ph}) &= \frac{\Delta Q_{ph}}{\Delta d_{ph}}+\frac{\Delta Q_{sl}}{\Delta d_{ph}} = \frac{Q_{ph}(d_{max}) - Q_{ph}(d_{min})- (d_{max} - d_{min})I_{sl}}{(d_{max} - d_{min})}
\end{align}
The sensitivity on the current $I_{ph}$ consumed during a phase $ph$ is 
\begin{align}
S(I_{ph}) &= \frac{\Delta Q_{ph}}{\Delta I_{ph}}  = \frac{Q_{ph}(I_{max}) - Q_{ph}(I_{min})}{I_{max} - I_{min}}
\end{align}
By assuming that the current $I_{ph}$ and the duration $d_{ph}$ of a phase are independent from each other, one can further simplify:
\begin{align}
S(I_{ph}) = d_{ph} ,& & S(d_{ph}) = I_{ph} - I_{sl}
\end{align}

In case of the duration of the post-processing phase $d_{post}$, which is the parameter having the strongest variations in our measurements, the change in energy-consumption of a connection interval is (using $S(d_{post})=I_{post}-I_{sl}=8.0mA$, $\Delta d_{post}=0.5ms$, $Q_{total}=25\mu C$):
\begin{align}
\Delta Q_{total}(d_{post}) = \Delta d_{post}\cdot S(d_{post})= 0.5ms\cdot 8mA = 4.0\mu C
\end{align}
The relative change $\frac{\Delta Q_{total}}{Q_{total}}$ is $16 \%$.
This means, due to variations in the post-processing phase, the energy consumption of a whole connection event might vary by $16 \%$.

The relative change in the energy consumption of a connection interval caused by variations in the transmission current $I_{tx}$ is (using $S(I_{tx})=d_{tx}\cdot N_{seq}=350\mu s$, $\Delta I_{tx}=10.0mA$, $Q_{total}=29\mu C$):
\begin{align}
\frac{\Delta Q_{total}}{Q_{total}} &= \frac{10 mA \cdot 350 \mu s}{29 \mu C}= 12\%
\end{align}
\label{sec:sensitivity}
\subsection{Using the model}
\label{sec:using_the_model}
With this paper, an implementation of our proposed energy model has been made available \cite{kindt:14}.
It is written as a C-library that provides easy-to-use functions for estimating the energy consumption of a BLE device and comes with a complete documentation in Doxygen format. It has a small resource demand and does not rely on any external code except some standard libraries. Therefore, it can easily be ported to embedded platforms.
The usage of the library is intuitive and illustrated with many examples. For example, to calculate the charge-consumption of one connection interval of $10$ ms with 5 pairs of packets per event, with $10$ bytes received and $20$ bytes sent per packet and with $3 dBm$ transmission power, the following code example can be used. 
\small \begin{verbatim}
#include "ble_model.h"
int main(){
	double charge;
	charge = ble_e_model_c_getChargeConnectionIntervalSamePayload(0,0.1,5,10,20,3);
return 0;
}
\end{verbatim}

\subsection{Designer Guidelines}
\label{sec:designer_guidelines}
In this section, we present possible uses of our model, leading to guidelines for system-designers on optimizing their parameter values for power consumption.
\subsubsection{Advertising/Scanning}
\begin{figure*}[htb]
\centering
\includegraphics[width = 13.5 cm]{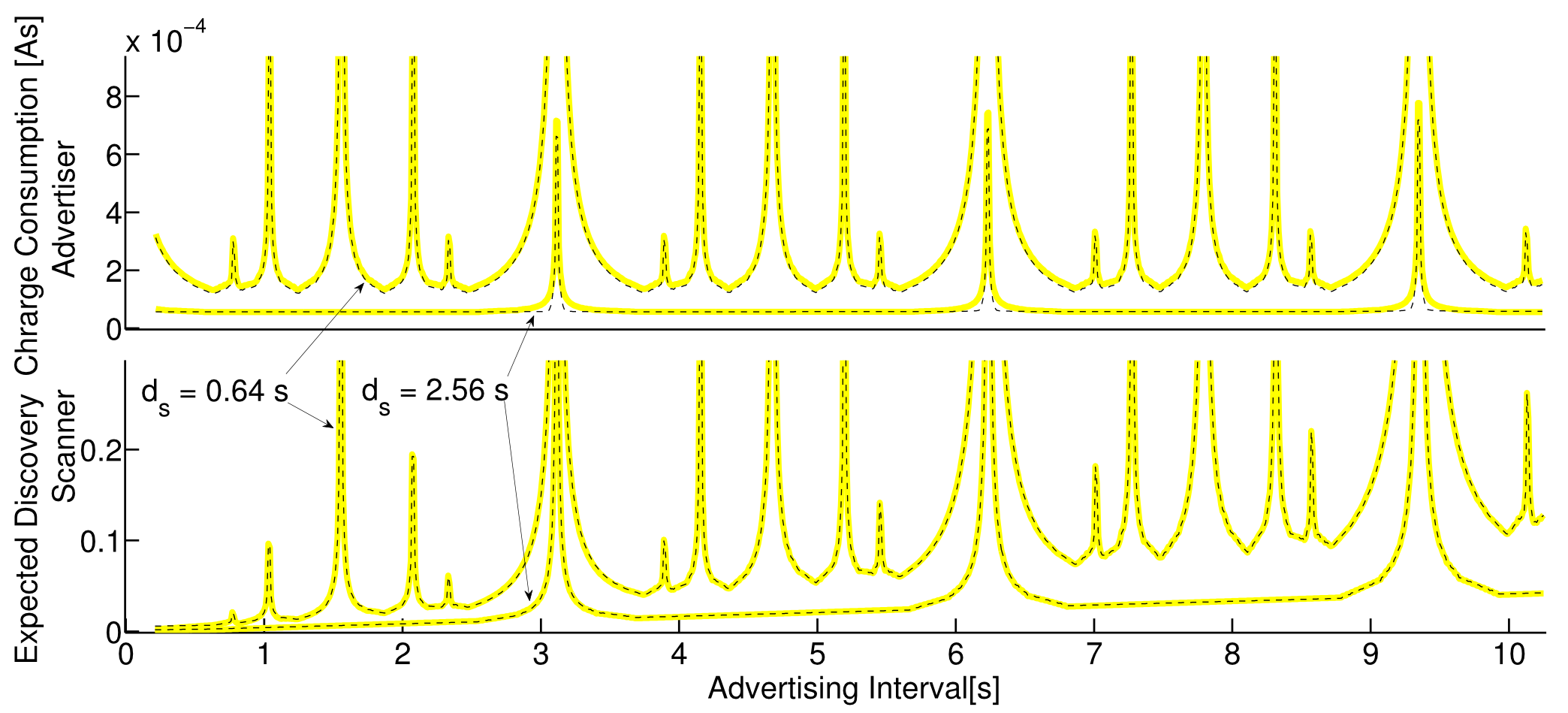}
\caption{Expected advertising/scanning charge for the following parameter values: $T_s = 3.12s$, $d_a = 446 \mu s$, $d_{ch} = 150 \mu s$, $\Delta = 93.12 ms$ , $\epsilon = 0.9999$. The charges consumed by the scan-events have been calculated using a simplified model, assuming that this charge is $i_{rx,S} \cdot d_{s}$ per event. The dashed lines show the approximated discovery charge computed using the discovery-latency plus the single-event model from Section \ref{sec:singleEventModel} only, whereas the solid lines show the more precise results for the charge calculated directly within Algorithm \ref{alg:advScanDur} using the additional assignment as already mentioned.}
\label{fig:expectedAdvScanEnergy_TaWs} 
\end{figure*}
Figure \ref{fig:expectedAdvScanEnergy_TaWs} shows the expected charge consumed by the advertiser and the scanner for different scan windows $d_s$. The solid lines depict the charges calculated directly within Algorithm \ref{alg:advScanDur}, whereas the dashed lines that overlap with the solid lines depcit the result using the Formulas \ref{eq:expectedAdvEnergy1}, \ref{eq:expectedAdvEnergy2}, \ref{eq:expectedAdvEnergy3} and \ref{eq:energyScanner}. As can be seen, the estimates given by these formulas do not differ significantly from the more exact curves that have been calculated directly by Algorithm \ref{alg:advScanDur}. As the formulas define a relation between $\overline{d_{adv}}$ and the charge consumed, a good estimation for the energy consumption of the advertiser and scanner can be made if $\overline{d_{adv}}$ is given.

Figure \ref{fig:expectedAdvScanEnergy_TaWs} reveals some interesting results. First, the energy consumption of the scanner is significantly higher than the energy consumption of the advertiser for all parameter values examined. Second, 
increasing the advertising interval does not increase the energy much if $T_a$ is not chosen to be within one of the peaks. However, a high $T_a$ reduces the 
energy consumption of idle-advertising significantly. This makes the mode $T_a > d_s'$ a good choice for neighbor discovery of devices that are expected to spend a long time in idle-scanning/idle-advertising because their remote nodes seldom come into range and if a higher discovery-latency is uncritical. 
Another interesting outcome is that for the scanner, the scan window does not influence its energy consumption much if $T_a$ is small. As $T_a$ increases, bigger scan-windows are better in terms of energy consumed by the scanner. 

From the model results, we propose the following choices to energy-optimize neighbor discovery procedures:
\begin{enumerate}
\item For devices with short expected idle-scanning durations (both devices begin the advertising/scanning process at approximately the same time) we suggest using continuous scanning for achieving the fastest possible discovery time and the smallest possible discovery energy. This mode is also beneficial if the scanner has much more energy than the advertiser (for example because it is connected to a grid), because the energy consumption of the advertiser is minimized.
Example: Two sensor nodes connect to exchange data once an hour at the same point in time. The clocks on both sides do not drift more than a few microseconds within that hour, therefore both devices become available at the same time.
\item For devices with intermediate idle scanning times or if a maximum discovery latency is required, a good trade-off is to set $T_a < d_s$ without using constant scanning. If the parameters on both devices can be chosen freely, energy can be "shifted" from one device to the other by lowering $d_s$ and lowering $T_a$ or vice-versa. 
Example: A data transfer between two sensors is manually initiated by a user by pushing a button on each node. Therefore, the user first switches on the scanner and five seconds later he activates the advertiser.
\item For devices that are expected to spend a lot of time with idle-scanning/idle advertising before they come in range, parameterizing the sender with $T_a > d_s$ is beneficial in terms of average power consumption both for the master and the slave, as it allows very low duty cycles on both sides while the discovery latency and discovery energy consumption  is only increased moderately. A maximum discovery latency cannot be guaranteed in this mode, but on the average the discovery latency can be low if the devices are parametrized well. However, to choose the parameter values carefully is of great importance to prevent the devices from going into a coupling phenomenon that leads to an extraordinarily long discovery latency and a high energy consumption.
Example: A device using the \emph{find-me profile} specified by the Bluetooth SIG \cite{BleFindMeProf}. This might be a keyfob that is lost and therefore is being looked for by another device like a smartphone. We consider the case where the keyfob is not in range because it is in the pocket of a person in another room. For instance, 30 minutes pass until the person enters the room and the advertiser meets the scanner.
In the find-me profile specified by Bluetooth SIG\cite{BleFindMeProf}, a parametrization with $T_a > d_s$ is used after $30 ms$ of lower duty-cycle scanning. $T_s$ is recommended to be $2.56s$ with $d_s = 11.25 ms$ ("Option 2"). In this profile, the advertising interval after $30ms$ of advertising is suggested to be between $1s$ and $2.5 s$ \cite{BleFindMeProf}. According to our model, $T_s = T_a = 2.5s$ would lead to high discovery energies and mean latencies as high as $560 s$ - as this constellation leads to a coupling phenomenon. Therefore, we propose a different value selection and recommend that further versions of the profile forbid such parameter combinations.

To avoid this situation, we propose to plot $\overline{d_{adv}}$ for the range of parameters that are considered and choose an appropriate local minimum.
\end{enumerate}

\ifACM
\else
\FloatBarrier
\fi
\subsubsection{Connected mode}
\begin{figure}[htb]
\centering
\includegraphics[width = 13cm]{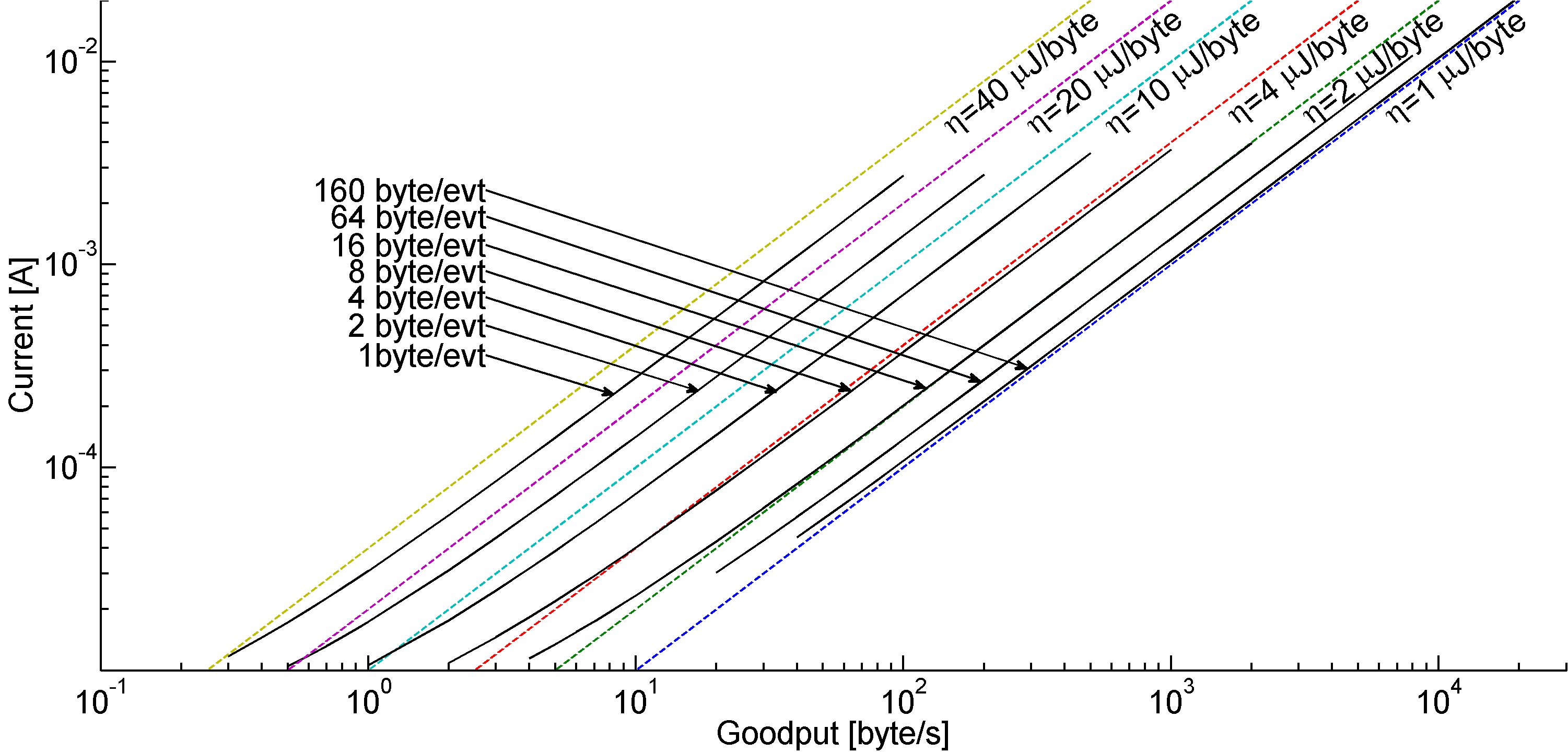}
\caption{Current consumption for different payload throughputs (goodput) using different payload sizes per event for a slave. The protocol overhead is 17 bytes per packet.}
\label{fig:goodput} 
\end{figure}
\begin{figure}[htb]
\centering
\includegraphics[width = 13cm]{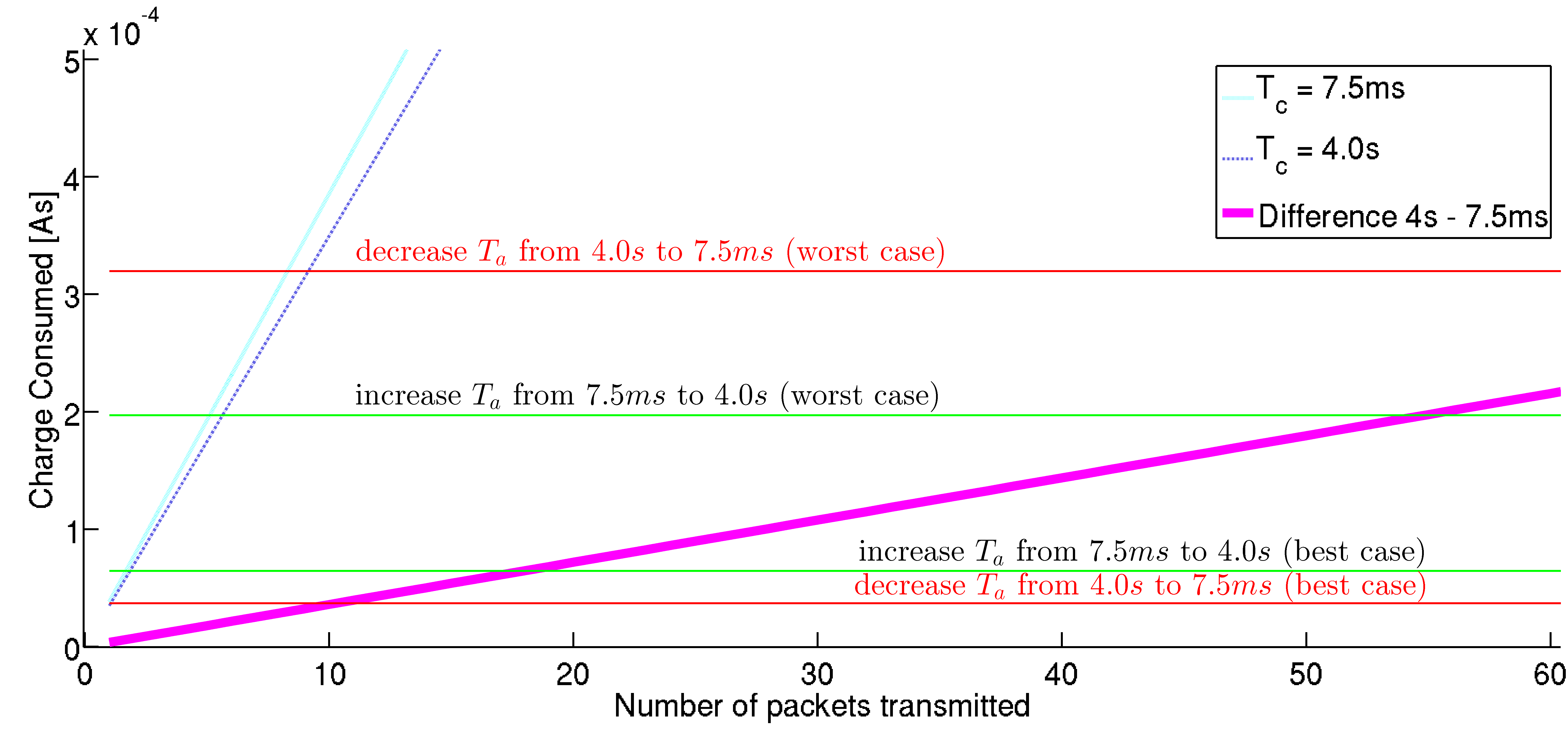}
\caption{Energy considerations when increasing or decreasing the connection interval for a slave. The horizontal lines depict the update costs. The bold line depicts the difference in charge consumed for both connection intervals.}
\label{fig:renegotiateTc} 
\end{figure}
\begin{figure}[htb]
\centering
\includegraphics[width = 13cm]{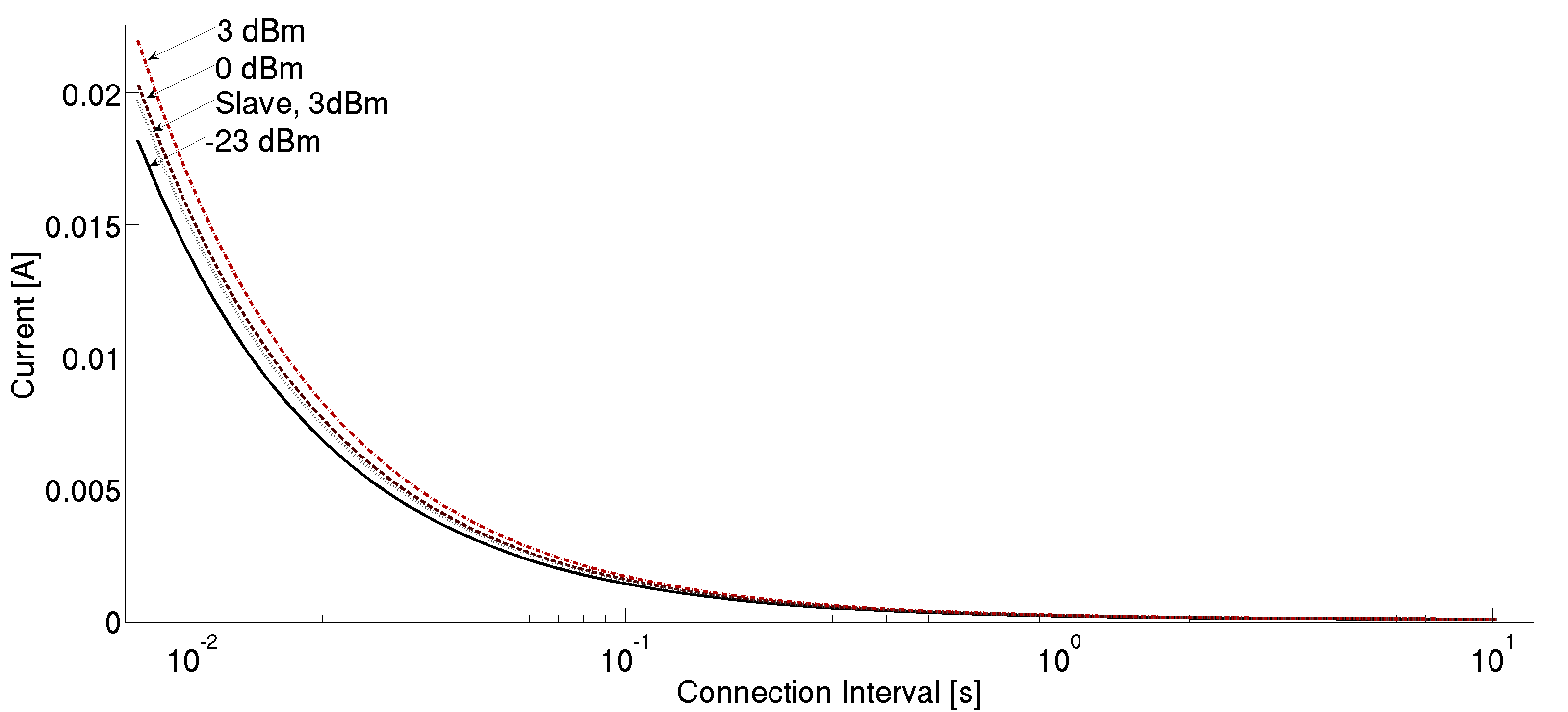}
\caption{Current consumption for different connection intervals and \emph{tx}-power levels for 8 packets per event with 20 bytes payload + 17 bytes overhead in each packet. The line for the slave shows the current the slave needs to receive the packets and respond with an empty packet.}
\label{fig:power_consumption_Tc_gtx} 
\end{figure}
For the connected mode, a parametrization needs to be chosen that optimizes the energy consumption for given constraints on minimum throughput and maximum latency.

Figure \ref{fig:goodput} shows the mean current of a BLE device with different amounts of payload bytes per connection event. Every packet sent contains 17 bytes protocol overhead and a maximum of 20 bytes payload. If there is more payload in an event than 20 bytes, two or more packets can be transmitted in one event. Given a fixed amount of payload per event,  different throughputs of the payload (goodputs) can be achieved by modifying the connection interval. In addition to the goodput-energy curves, different energy-efficiencies per byte are shown as diagonal lines in the figure. As can be seen, the transmission takes place with different efficiencies $\eta = \frac{bytes}{charge}$ for different parameter values. A parametrization drawing a higher current consumption can have a better efficiency per byte payload than others drawing a lower current. For example, the efficiency of the point with the smallest current and smallest goodput at the lower left of the figure is by far less than achieving the same throughput with 16 bytes per event. Another fact that can be observed is that for any of the curves with a fixed payload per event, the increase of the goodput leads to a rise of the efficiency but at the same time to an increased current consumption. In conclusion, different goodput values with different efficiencies can be achieved by choosing the number of packets per event, the number of payload bytes within each packet and the connection interval.

To achive a high efficiency $\eta$ for a given goodput, our suggestions are as follows.
\begin{enumerate}


\item Add as much payload in a packet as possible \cite{siekkinen:12}. In an ATT\_HANDLE\_NOTIFY-packet which is typically used for transmitting payload via ATT, the overhead is 17 bytes whereas the maximum payload per request is 20 bytes. Not utilizing the whole 20 bytes payload leads to a larger overhead per payload byte and to an additional overhead caused by the influential factors of the energy consumption that are independent from the payload length.
\item Maximize the number of (filled) packets per event \cite{siekkinen:12} while increasing the connection interval to the maximum value possible that still meets the throughput constraints required. In addition, Suggestion (1) should be kept and empty or sparsely filled packets should be avoided. As the connection interval is increased, a higher latency (time between data is ready on the sender and received by the receiver) occurs.
\end{enumerate}

A slightly different objective of optimization is a low total current consumption rather than a high efficiency for a fixed amount of payload to be sent. For example, this applies when data from a sensor is streamed with a fixed or a varying rate. For situations like this, we further suggest the following.
\begin{enumerate}
\setcounter{enumi}{2}\item When idle, the connection interval should be as long as possible to avoid empty polling packets, constrained by the maximum latency. Figure \ref{fig:power_consumption_Tc_gtx} shows the power consumption
for transmitting eight 20-byte-payload-packets per event for different connection intervals. As can be seen easily, a lot of power can be wasted by choosing an unnecessarily high connection interval. $T_c$ is the most important factor for the power consumption in connected mode and it is crucial to set it carefully. A higher power consumption for small connection intervals has also been measured in \cite{goemz:12}.
\item Figure \ref{fig:renegotiateTc} shows the charge consumed by a slave for transferring a variable amount of 20-byte-packets from the master to the slave with $T_c = 0.0075 s$ and $T_c = 5.0 s$. The bold line in this figure represents the difference in the charge consumption between the two connection intervals and is therefore equal to the savings if one would change $T_c$ from $7.5 ms$ to $4.0 s$. The horizontal lines depict the charge consumed for switching from $T_c = 7.5 ms$ to $T_c = 4.0 s$ and vice versa, according to the procedure described in Section \ref{sec:conProc}. As mentioned before, the energy consumption for this procedure depends on parameter values chosen by the stack and a best case/worst case estimation of the energy consumption can be made. We now consider after how many connection intervals increasing the $T_c$ from $7.5 ms$ to $4.0 s$ pays of energetically. In this example, in the best case, increasing the connection interval is beneficial after approximately 18 events. In the worst case, it is beneficial after approximately 55 events. The cost for switching back is also shown in the figure. The worst case charge for a connection update is much higher than for the best case due to the long possible transmit window $d_{tw}$, but should occur very rarely as the master normally uses small transmit window sizes and schedules the anchor point for the new interval reasonably. With these results, one should consider renegotiating the connection interval whenever there is a change in the throughput required. Our proposed model can be used to estimate if the expected savings in the energy consumption with the new connection interval prevails the cost for updating the connection parameters.
\item Using slave latency instead of lowering the connection interval is not beneficial if the throughput required by the slave is known, as the energy of the master rarely lowers with higher slave latency because it cannot skip the connection events. However, the curve in Figure \ref{fig:power_consumption_Tc_gtx} could be extended towards longer effective connection intervals for the slave by using slave latency. Therefore, there is a trade-off between renegotiating the connection interval frequently and keeping the connection interval constant while using slave latency: The slave might set a long effective connection interval if it is likely that there is few data to be sent, or it might just skip some connection intervals and save the power for renegotiation of the connection interval if the data rate is likely to rise again. The latter will result in a higher energy consumption for the master as, opposed to the slave, it cannot skip any connection interval.
\item Figure \ref{fig:power_consumption_Tc_gtx} reveals that energy can be saved if the \emph{tx}-power-level is adjusted to a value that is sufficient to transmit the data across the distance between sender and receiver without loss instead of sending with the maximum power.
The curves for different \emph{tx}-powers lie in close proximity. Whereas it is desirable to send with the lowest possible power the device and the application allows, the impact of this factor is much smaller than others such as the connection interval.

\end{enumerate}
\ifACM
\else
\FloatBarrier
\fi

\subsection{Limitations}
\label{sec:limitations}
This paper described methods for estimating the energy consumption of a BLE device. 
However, there are some limitations.
\subsubsection{Exponential peaks in advertising/scanning}
As we abort Algorithm \ref{alg:advScanDur} for all times $d_{exp} > d_{exp,max}$, we cannot model the center of exponential peaks. In close vicinity of these peaks, we loose accuracy. 
As the model is capable of predicting these peaks, the discovery-latencies and -energies in these peaks are not of interest as no designer would choose such a parametrization. Only the exact values of those parameters that are not within these peaks matter. 
\subsubsection{Ambiguous model parameters in advertising/scanning}
Within Algorithm \ref{alg:advScanDur}, there are two parameters $\epsilon$ and $\Delta$ that affect the quality of the model results. At the moment, we cannot provide a meta-model that gives an estimation of the error depending on the quality parameters. Setting them unnecessarily high will lead to a long runtime of the model algorithm. Until now, these values have to be chosen based on a good guess.
\subsubsection{Repetated Broadcasting}
We presented a model for the average advertising latency $\overline{d_{adv}}$. Therefore, we can model a pair of nodes discovering each other for connection establishment and we can model the average latency until reception if a device periodically sends data in non-connected mode and changes the data after random amounts of time (for example when a sensed value changes).
However, there is a case that cannot be modeled properly, yet: Consider the situation where one node is scanning and the other node is broadcasting data repeatedly, without getting an acknowledgement when data is received. The same broadcast data is repeated with $T_a$ until new data is ready. Now we can calculate the average latency $\overline{d_{adv}}$ as described, which is the average duration until the first successful reception takes place. For the second reception, our scheme cannot be applied as the new offset $\phi$ is not uniformly distributed anymore. It depends on the first offset, on the number of advertising events that occurred and on the accumulated sum of random advertising delays $\sum\limits_{1}^{n} \rho$.
\subsubsection{Preprocessing/Postprocessing}
The preprocessing and postprocessing durations vary from event to event. As it significantly contributes to the overall power consumption, an exact model that is capable of predicting the actual values in each event would be preferable over current models that can only predict mean values. The details needed to model the processing durations precisely are hidden within the firmwares of the BLE modules - it falls into the responsibility of the firmware developers to make this data available.

\subsection{Concluding remarks}
In this paper, a precise and comprehensive model of the BLE protocol was presented. The model parameter values for BLE112-devices were given. These model-parameters need to be remeasured for other devices. The model has proven good accuracies - in a test-scenario, an error of the predicted current of less than $6 \%$ in connected mode was measured. The accuracy can be increased by using individual sets of model-parameters for different BLE functions. The proposed model can be used for battery lifetime calculations, for predictions on worst case peak currents and for deriving power management strategies. With the results presented, power-management-algorithms for updating the connection parameters adaptively to the current situation can be developed. 

Further research on an error estimation for the parameters $\Delta$ and $\epsilon$ of Algorithm \ref{alg:advScanDur} seems to be desirable for reducing the runtime of the algorithm. An open research topic is the reception probability in continuous broadcasting scenarios. In addition, refined models for the pre- and postprocessing durations should be derived in the future for further increasing the model's accuracy.

Modeled results have been compared to measured ones for the connected mode and for the neighbor discovery latency. Further comparative measurements are required to determine the accuracy of the model for scan events and for the connection establishment/update procedure model. However, as the models have been derived from real world measurements, the expected accuracies should be similar to the ones for the connected mode.

\section{Credits}
This work was partially supported by "HE2mT - High-Level Development Methods for Energy-Saving, Mobile Telemonitoring Systems", a project funded by the Federal Ministry of Education and Research of Germany. 
Author's emails: \{kindt, yunge, diemer, chakraborty\}@rcs.ei.tum.de
Author's addresses: Lehrstuhl f\"ur Realzeit-Computersysteme, Technische Universit\"at M\"unchen, Arcisstr. 21, 80333 M\"unchen, Germany.
This paper is in submission to a journal. If it is accepted, it will be updated such that a link to the published paper and the DOI will be provided. This document contains the non-reviewed version of the authors.
\bibliographystyle{plain}
\bibliography{bibliography_acm}

\begin{appendix}
\section{Parameter values for BLE112 in connected mode}
\label{sec:appendixParamsConnected}
In the two tables of this section, we present values for all model parameters of the connected mode for a BLE112-device. The slave clock latency is reported to be between $31$ and $50\mbox{ } ppm$ by the device during the handshaking for the connection establishment process. In this paper, we therefore calculate with a sleep clock accuracy of $50 ppm$. The duration $d_{prerx}$ is longer than its table value for the first \emph{rx}-phase within a connection event of a slave. For the latter case, $d_{prerx}$ is $388 \mu s$.
The values presented have been measured by analyzing about 4000 connection events of a BLE-attribute write procedure without confirmation of the remote node for different payload lengths and different numbers of events per packet. The connection interval $T_c$ was $100 ms$, the \emph{tx}-power was $3 dBm$. We also performed the measurement for different \emph{tx}-power levels. These \emph{tx}-power levels mainly result in different currents for the \emph{tx}-section. Other parameters such as $Q_{cso}$ and $d_{tra}$ are influenced too, but the impact on their values and hence on the event energy consumption is much lower than the impact of the \emph{tx}-current. For a good approximation, we assume all parameters except the \emph{tx}-current to constant for different tx-power levels in our model. In this paper, we calculated with a sleep current $I_{sl}$ of $0.9 \mu A$ according to \cite{ble112}.

\ifACM
\begin{acmtable}{14cm}[h]
\else
\begin{table}[h]
\fi
\centering

\begin{tabular}{|c|c|c|c|c|c|c|c|c|}
\hline  Phase & $d_{avg} $ & $d_{min}$, & $d_{max}$ & $d_{std}$ & $I_{avg}$ & $I_{min}$  & $I_{max}$ & $I_{std}$ \\ 
			& [ms] & [ms] & [ms] & [ms] & [mA] & [mA] & [mA] & [mA] \\
\hline
\hline  head&0.578&0.500&0.640&0.012&5.924&5.558&6.165&0.085 \\ 
\hline  pre&0.305&0.010&0.450&0.109&7.691&5.570&7.997&0.153 \\ 
\hline  rx&-&-&-&-&26.505&25.967&27.676&0.302\\ 
\hline  rxtx&0.080&0.060&0.100&0.004&14.128&13.793&14.653&0.115\\ 
\hline  tx&-&-&-&-&36.445&35.571&38.763&0.559\\ 
\hline  pretx&0.053&0.014&0.084&0.018&-&-&-&-\\ 
\hline  txrx&0.057&0.040&0.070&0.005&15.125&14.605&16.048&0.198\\ 
\hline  cpre&0.073&0.050&0.080&0.004&12.238&11.633&13.006&0.200\\ 
\hline  prerx&0.123&0.110&0.140&0.005&-&-&-&-\\ 
\hline  tra&0.066&0.040&0.090&0.011&11.636&8.964&14.721&1.416\\
\hline  post&0.860&0.610&1.110&0.101&7.980&7.919&8.221&0.065\\ 
\hline  tail&0.080&0.060&0.340&0.013&4.129&3.088&6.995&0.380 \\ 
\hline 
\hline Phase&$Q_{avg}$&$Q_{min}$&$Q_{max} $&$Q_{std}$&-&-&-&-\\
& [uC] & [uC] & [uC] & [uC] &  &  &  &  \\
\hline to& -1.2&-1.8&-0.8&0.2&-&-&-&-\\ 
\hline 
\end{tabular}
\caption{Model parameteres for connected mode with a tx-power of $3 dBm$ for a BLE112 transceiver. Average ($X_{avg}$), minimal($X_{min}$) and maximal ($X_{max}$) values are given along with the standard deviation $X_{std}$.}

\ifACM
\end{acmtable}
\else
\end{table}
\fi

\ifACM
\begin{acmtable}{14cm}[h]
\else
\begin{table}[h]
\fi
\centering
\begin{tabular}{|l|c|}
\hline
\emph{tx}-power & $I_{tx} [mA]$\\
\hline
\hline
$3 dBm$&$36.5$\\
\hline
$2 dBm$&$33.5$\\
\hline
$0 dBm$&$32.1$\\
\hline
$-1 dBm$&$31.5$\\
\hline
$-2 dBm$&$30.6$\\
\hline
$-3 dBm$&$30.1$\\
\hline
$-5 dBm$&$29.1$\\
\hline
$-6 dBm$&$28.8$\\
\hline
$-8 dBm$&$28.4$\\
\hline
$-10 dBm$&$28.1$\\
\hline
$-12 dBm$&$27.9$\\
\hline
$-15 dBm$&$27.7$\\
\hline
$-17 dBm$&$27.6$\\
\hline
$-19 dBm$&$27.5$\\
\hline
$-21 dBm$&$27.5$\\
\hline
$-23 dBm$&$26.3$\\
\hline
\end{tabular}
\caption{Current of the transmission-phase for all supported tx-power-levels of a BLE112-transciever.}

\ifACM
\end{acmtable}
\else
\end{table}
\fi

\FloatBarrier
\section{Parameter values for BLE112 in scanning mode}
The table below presents values for all model parameters for the scanning mode. They were obtained by analyzing 1257 scan events for a 
tx-power of $3 dBm$.
\ifACM
\begin{acmtable}{14cm}[h!]
\else
\begin{table}[h]
\fi
\centering
\label{table:dataForScanning}
\begin{tabular}{|c|c|c|c|c|c|c|c|c|}
\hline  Part & $d_{avg} $ & $d_{min}$, & $d_{max}$ & $d_{std}$ & $I_{avg}$ & $I_{min}$  & $I_{max}$ & $I_{std}$ \\ 
			& [ms] & [ms] & [ms] & [$\mu s$] & [mA] & [mA] & [mA] & [mA] \\
\hline
\hline  pre,s&0.700&0.680&0.730&10&7.087&6.924&7.253&0.065 \\ 
\hline  rx,s&-&-&-&-&26.399&26.042&26.480&0.043\\ 
\hline  rxtx,s&0.115&0.110&0.120&0.498&15.011&14.617&15.519&0.288\\ 
\hline  tx,s&-&-&-&-&35.999&35.650&36.488&0.247\\ 
\hline  pretx,s&0.014&4e-3&0.024&1.184&-&-&-&-\\ 
\hline  txrx,s&0.089&0.080&0.090&2.332&16.670&15.875&17.224&0.244\\ 
\hline  rxsr&-&-&-&-&26.426&26.279&26.563&0.058 \\ 
\hline  prerx,s&0.074&0.068&0.088&2.45&-&-&-&-\\ 
\hline  rxrx,s &0.377&0.370&0.380&4.488&9.633&9.426&9.768&0.11\\ 
\hline  post,s&0.816&0.710&1.820&246&8.012&7.820&8.138&0.060\\ 
\hline  chch,s&1.325&1.320&1.330&4.983&8.550&8.470&8.624&0.042\\ 
\hline 
\hline Part&$Q_{avg}$&$Q_{min}$&$Q_{max} $&$Q_{std}$&-&-&-&-\\
& [uC] & [uC] & [uC] & [uC] &  &  &  &  \\
\hline ctx,s& -0.2264&-0.3244&-0.1456&0.0143&-&-&-&-\\ 
\hline crx,s& -0.1350&-0.1900&-0.0851&0.0123&-&-&-&-\\ 
\hline 
\end{tabular}
\caption{Model parameter values for scan events. Average ($X_{avg}$), minimal($X_{min}$) and maximal ($X_{max}$) values are given along with the standard deviation $X_{std}$.}

\ifACM
\end{acmtable}
\else
\end{table}
\fi

\FloatBarrier
\section{Table of symbols}
The symbol $X$ stands either for the charge $Q$, the current $I$ or the duration $d$ of a phase of a connection event.
\begin{longtable}{|p{2.0cm}|p{1.5cm}|p{8.0cm}|}
\hline \multicolumn{3}{|c|}{\textbf{BLE Protocol}}\\
\hline  \textbf{Symbol} & \textbf{Unit} & \textbf{Description} \\ 
\hline  $T_a$& [s] & Advertising interval \\ 
\hline  $T_s$ & [s] & Scan interval \\ 
\hline  $d_s$ & [s] & Scan window  \\ 
\hline  $d_{IFS}$ & [s]  & Interframe-space \\ 
\hline  $\rho(t) $& [s] & Random advertising delay \\ 
\hline  $d_{tw}$ & [s]  & Transmit window  \\ 
\hline  $d_{two}$ & [s]  & Transmit window offset \\ 
\hline  $t_{anchor}$& [s]  & Anchor point: Reference point for connection interval  \\ 
\hline $T_c$ & [s] & Connection interval \\ 
\hline $T_{c,o}$ & [s]  & Connection interval before a connection parameter update\\ 
\hline $T_{c,n}$ & [s] &  Connection interval after a connection parameter update\\ 
\hline $d_{c}$ & [s] & Duration of a connection event  \\ 
\hline $N_{sl}$ & - & Slave latency  \\ 
\hline $\overline{d_{adv}}$ & [s] & Expected discovery latency in the advertising/scanning mode  \\ 
\hline \multicolumn{3}{|c|}{\textbf{Single event model for the connected mode}}\\
\hline  \textbf{Symbol} & \textbf{Unit} & \textbf{Description} \\ 
\hline  $X_{head}$&[C/A/s]& Device wakeup phase\\ 
\hline  $X_{pre}$ &[C/A/s]& Preprocessing  phase\\ 
\hline  $X_{cpre}$&[C/A/s]& Communication preamble phase\\ 
\hline  $X_{ww}$&[C/A/s]& Window widening phase\\ 
\hline  $X_{prerx}$&[C/A/s]& Constant part of reception phase\\ 
\hline  $X_{rx}$&[C/A/s]& Reception phase\\ 
\hline  $N_{rx}$&-& Number of bytes received\\ 
\hline  $X_{IFS}$ &[C/A/s]& Interframe space  \\ 
\hline  $X_{rxtx}$&[C/A/s]& Reception-to-transmission transition phase\\ 
\hline  $X_{txrx}$&[C/A/s]& Transmission-to-reception transition phase\\ 
\hline  $X_{pretx}$&[C/A/s]& Constant part of transmission phase\\ 
\hline  $X_{tx}$ &[C/A/s]& Transmission phase\\ 
\hline  $N_{tx}$&-& Number of bytes sent\\
\hline  $X_{tra}$ &[C/A/s]& Transient phase\\ 
\hline  $X_{post}$ &[C/A/s]& Postprocessing phase\\ 
\hline  $X_{tail}$ &[C/A/s]& Tail phase (i.e., phase that initiates of the sleep mode)\\ 
\hline  $d_{seq}$ &[s]& Duration of a communications sequence (i.e., exchange of one pair of packets)\\ 
\hline  $d_{event}$ &[s]& Duration of a whole connection event (from head to tail)\\ 
\hline  $X_{t}$ &[C/A/s]& Communication sequence (reception, transmission, interframe-spaces and correction term for capacitances/resistances)\\ 
\hline  $Q_{to}$ &[C]& Correction term for capacitances/resistances in power supply line\\ 
\hline  $SCA_{Ma/Sl}$ &[ppm]& Sleep clock accuracy of master/slave\\ 
\hline  $I_{sl}$&[A]& Sleep current of the BLE device \\ 
\hline 
\hline \multicolumn{3}{|c|}{\textbf{Single event model for scan events}}\\
\hline  \textbf{Symbol} & \textbf{Unit} & \textbf{Description} \\ 
\hline  $X_{pre}$&[C/A/s]& Wakeup and preprocessing phase\\ 
\hline  $X_{S,1}$&[C/A/s]& Phase of scanning until the first reception takes place\\ 
\hline  $X_{rx,s}$&[C/A/s]& Reception phase (advertising packet is received)\\
\hline  $X_{rxtx,s}$&[C/A/s]& Reception-to-transmission transition phase\\ 
\hline  $X_{tx,s}$&[C/A/s]& Transmission of scan-request (or connection request) phase\\ 
\hline  $X_{txrx,s}$&[C/A/s]& Transmission-to-reception transition phase\\ 
\hline  $X_{rxs}$&[C/A/s]& Phase of the reception of a scan response\\ 
\hline  $X_{rxrx,s}$&[C/A/s]& Reception-to-scanning transition phase\\
\hline  $X_{S,2}$&[C/A/s]& Phase of scanning until the end of the scan event\\ 
\hline  $X_{post,s}$&[C/A/s]& Postprocessing phase \\ 
\hline  $I_{scan}$&[A]& Current consumption of the device while scanning \\ 
\hline  $Q_{sEv}$&[C]& Charge consumed by a scan event (general case) \\ 
\hline  $Q_{sEv,Idle}$& [C] & Charge consumed by a scan event (idle scanning) \\ 
\hline  $Q_{crx,s}$& [C] & Correction term for non-rectangular shapes of the reception-phase(s)\\ 
\hline  $Q_{ctx,s}$& [C] & Correction term for non-rectangular shapes of the transmission-phase(s)\\ 
\hline 
\hline \multicolumn{3}{|c|}{\textbf{Connection procedure model}}\\
\hline  \textbf{Symbol} & \textbf{Unit} & \textbf{Description} \\ 
\hline  $Q_{ev,cR,Ma/Sl}$& [C]& Charge consumed by the master/slave for an event a connection request event is sent within\\ 
\hline  $Q_{ev,cU,Ma/Sl}$ &[C] & Charge consumed by the master/slave for an event a connection update packet is sent within\\ 
\hline  $d_{sl,cR/cU}$ & [s]& Sleep duration after a connection request/update event\\ 
\hline  $Q_{cE,Ma/Sl}$ & [C]& Charge consumed for a connection establishment by the master/slave \\ 
\hline  $Q_{cU,Ma/Sl}$ & [C]& Charge consumed for a connection parameter update by master/slave\\ 
\hline  $d_p$ & [s]& Time from beginning of the transmit window until the master sends the first packet with new $T_c$\\ 
\hline  $d_{ww, cE/cU}$ & [s]& Window widening for connection establishment/parameter update \\ 
\hline 
\hline \multicolumn{3}{|c|}{\textbf{Discovery Latency Model}}\\
\hline  \textbf{Symbol} & \textbf{Unit} & \textbf{Description} \\ 
\hline  $d_{advPkg}$ &[s]&Duration of an advertising packet\\ 
\hline  $\phi$ &[s]& Offset between the beginning of the first scan event and the first advertising event \\ 
\hline  $T_{a,0}$ &[s]& Constant part of the advertising interval \\ 
\hline  $\rho$ &[s]& Random delay for advertising events\\ 
\hline  $t_{i,suc}$&[s]&Set of points in time an advertising packet which is sent within is received successfully\\ 
\hline  $d_{early}$ & [s]&Time an advertising event can be sent before the beginning of the corresponding scan event to be received successfully\\ 
\hline  $d_{late}$ & [s]&Time an advertising event must be sent before the end of the corresponding scan event to be received successfully\\ 
\hline  $d_{s}'$ & [s]& Effective scan window ($d_s' = d_s - d_a$) \\ 
\hline  $d_{a}$ & [s]&Duration of an advertising packet\\ 
\hline  $d_{ch}$ & [s] &Time the advertiser needs for changing the channel it sends its packets on\\ 
\hline  $t_{sE}$&[s]&Time a scan event starts at\\ 
\hline  $t_{advEvt}$&[s]&Time an advertising event is sent at\\ 
\hline  $p_{hit}(n)$&-&Probability of an advertising event $n$ for hitting a scan event \\ 
\hline  $f(\rho)$&-&Probability density function for the random advertising delay\\ 
\hline  $\mu$&-&Expected value\\ 
\hline  $\sigma$&-&Standard deviation\\ 
\hline  $f(t_{advEvt})$&-&Probability density function for the beginning of an advertising event\\ 
\hline  $p_k$&-&Probability for an advertising event being received in the scan event $k$\\ 
\hline  $\Phi(t)$&-&Standard normal cumulative distribution \\ 
\hline  $t_{ai}$&[s]&Point in time an advertising event would take place without taking the random delay $\rho$ into account\\ 
\hline  $k_{min}(n)$, $k_{max}(n)$&-&Lowest and highest index of a scan event the advertising event $n$ could overlap with\\ 
\hline  $p_{cM}(n)$&-&Cumulative miss probability (probability that n advertising events in a row miss the scan event)\\ 
\hline  $d_{exp}$&[s]&Expected advertising duration for a given pair of values $\Phi$ and $k$\\ 
\hline  $\epsilon$&-&Parameter of Algorithm \ref{alg:advScanDur} that determines the preciseness and computational complexity\\ 
\hline  $ch$&-&Channel number (37/38/39)\\ 
\hline  $d_{exp,max}$&[s]&Maximum expected advertising duration for a given pair of values $\phi$ and $k$ before the calculation is aborted\\ 
\hline  $\Delta$&[s]&Interval between two offsets $\phi$ that are examined\\ 
\hline  $\overline{d_{adv}}$&[s]&Expected discovery latency\\ 
\hline  $P_{37}/P_{38}/P_{39}$&-&Reception probability on channel 37/38/39 in the constant scanning mode\\
\hline  $P_{loss}$&-&Probability that the first advertising packet is lost in the constant scanning mode\\ 
\hline  $d_{adv,max}$&[s]&Maximum possible discovery latency in the constant scanning mode\\ 
\hline  $h_1$, $h_2$, $h_3$, $A$, $B$, $C$&var.&Helper functions for constant scanning model\\ 
\hline 
\hline \multicolumn{3}{|c|}{\textbf{Estimating the number of events}}\\
\hline  \textbf{Symbol} & \textbf{Unit} & \textbf{Description} \\ 
\hline  $N_{c,ma/sl}$&-&Number of connection events of the master/slave within a given amount of time\\ 
\hline  $\overline{Q_{con}}$&[C]&Overall charge consumed by the master/slave in the connected mode\\ 
\hline  $T_{g}$&[s]&Given amount of time which is considered\\ 
\hline  $\overline{N_{sl}}$&-&Average slave latency\\ 
\hline  $X_{event}$&[C/A/s]&Charge/current/duration of an arbitrary event\\ 
\hline  $X_{full}$&[C/A/s]&Full advertising event\\ 
\hline  $X_{last}$&[C/A/s]&Last advertising event\\ 
\hline  $\overline{Q_{adv}}$&[C]&Expected charge consumed by an advertiser (whole discovery process)\\ 
\hline  $Q_{advEvent}$&[C]&Charge consumed by an advertising event\\ 
\hline  $X_{37(38/39)}$&[C/A/s]&Advertising event on channel 37/38/39\\ 
\hline  $\overline{N_a}$&-&Expected number of advertising events\\ 
\hline  $\overline{Q_{s}}$&[C]&Expected Energy of scanner (whole discovery process)\\ 
\hline  $\overline{Q_{active}}$&[C]&Expected charge of scanner for all times it is not sleeping\\ 
\hline  $\overline{Q_{sleep}}$&[C]&Expected charge of scanner for all times it is sleeping\\ 
\hline  $Q_{sEv,idle}$&[C]&Charge consumed by an idle scan event (i.e., scan event without reception)\\ 
\hline  $\overline{N_a}, \overline{N_s}$&[C]&Expected number of advertising/scan events during neighbor discovery\\ 
\hline 
\hline \multicolumn{3}{|c|}{\textbf{Model validation and sensitivity analysis}}\\
\hline  \textbf{Symbol} & \textbf{Unit} & \textbf{Description} \\ 
\hline  $\varepsilon_{m}$&-&Relative error between measured and modeled values\\ 
\hline  $S$&var.&Sensitivity\\ 
\hline  $Q_{ph}$&[C]&Charge consumed by a phase/state the device is in (e.g., preprocessing, reception,..)\\ 
\hline  $d_{min}/d_{max}$&[s]&Minimum/maximum phase duration\\ 
\hline  $I_{min}/I_{max}$&[A]&Minimum/maximum current of a phase\\ 
\hline  $\delta Q_{total}$&[C]&Variation of the total charge consumed in a given phase\\ 
\hline  $\eta$&[Bytes/C]&Efficiency\\ 
\hline 
\end{longtable}

\end{appendix}

\end{document}